\documentclass[12pt]{article}
\usepackage{amsmath,amssymb}
\usepackage{graphicx}
\newcommand{\eqdef}{\stackrel{\text{def}}{=}}
\newcommand{\eqdefrm}{\stackrel{\text{\rm def}}{=}}
\newcommand{\n}{\nonumber\\}
\newcommand{\bm}{\boldsymbol}
\newcommand{\ignore}[1]{}
\numberwithin{equation}{section}
\newcommand{\Romannumeral}[1]{\uppercase\expandafter{\romannumeral#1}}

\newcommand{\II}{\text{\Romannumeral{2}}}
\newcommand{\III}{\text{\Romannumeral{3}}}
\newcommand{\IV}{\text{\Romannumeral{4}}}
\newcommand{\V}{\text{\Romannumeral{5}}}
\newtheorem{thm}{\bf Theorem}
\newtheorem{conj}{\bf Conjecture}

\allowdisplaybreaks[4]

\begin{document}

\baselineskip=20pt

\newfont{\elevenmib}{cmmib10 scaled\magstep1}
\newcommand{\preprint}{
    \begin{flushright}\normalsize \sf
     DPSU-18-2\\
   \end{flushright}}
\newcommand{\Title}[1]{{\baselineskip=26pt
   \begin{center} \Large \bf #1 \\ \ \\ \end{center}}}
\newcommand{\Author}{\begin{center}
   \large \bf Satoru Odake \end{center}}
\newcommand{\Address}{\begin{center}
     Faculty of Science, Shinshu University,\\
     Matsumoto 390-8621, Japan
   \end{center}}
\newcommand{\Accepted}[1]{\begin{center}
   {\large \sf #1}\\ \vspace{1mm}{\small \sf Accepted for Publication}
   \end{center}}

\preprint
\thispagestyle{empty}

\Title{Dual Polynomials of the Multi-Indexed ($q$-)Racah Orthogonal
Polynomials}

\Author

\Address
\vspace{1cm}

\begin{abstract}
We consider dual polynomials of the multi-indexed ($q$-)Racah orthogonal
polynomials. The $M$-indexed ($q$-)Racah polynomials satisfy the second
order difference equations and various $1+2L$ ($L\geq M+1$) term recurrence
relations with constant coefficients.
Therefore their dual polynomials satisfy the three term recurrence relations
and various $2L$-th order difference equations.
This means that the dual multi-indexed ($q$-)Racah polynomials are ordinary
orthogonal polynomials and the Krall-type.
We obtain new exactly solvable discrete quantum mechanics with real shifts,
whose eigenvectors are described by the dual multi-indexed ($q$-)Racah
polynomials. These quantum systems satisfy the closure relations, from which
the creation/annihilation operators are obtained, but they are not shape
invariant.
\end{abstract}

\section{Introduction}
\label{sec:intro}

Ordinary orthogonal polynomials in one variable satisfying second order
differential equations are severely restricted by Bochner's theorem
\cite{bochner,ismail}. Allowed polynomials are the Hermite, Laguerre, Jacobi
and Bessel polynomials, but the weight function of the Bessel polynomial is
not positive definite.
Various attempts to avoid this no-go theorem have been carried out and there
are three directions.

The first direction (\romannumeral1) is to change the second order to higher
orders. This direction was initiated by Krall \cite{k38} and he classified
the orthogonal polynomials satisfying fourth order differential equations
\cite{k40}.
Based on the Laguerre and Jacobi polynomials, by adding the Dirac delta
functions to the weight functions, orthogonal polynomials satisfying
higher order differential equations are obtained
\cite{l82}--\cite{z99}.
Such polynomials are called the Krall polynomials.
The second direction (\romannumeral2) is to replace a differential equation
with a difference equation.
Studies in this direction were summarized as the Askey scheme of
(basic-)hypergeometric orthogonal polynomials and various generalizations of
the Bochner's theorem were proposed \cite{kls,ismail}.
By combining (\romannumeral1) and (\romannumeral2), orthogonal polynomials
satisfying higher order difference equations were also studied.
We call such polynomials the Krall-type polynomials.
Some of them have weight functions with delta functions
\cite{vz01}
and some others have those without delta functions
\cite{d12}--\cite{ad13}.
The third direction (\romannumeral3) is to allow missing degrees.
This means the following situation: polynomials
$\{\mathcal{P}_n\}$ ($n\in\mathbb{Z}_{\geq 0}$) are orthogonal and
satisfy second order differential equation and form a complete set,
but there are missing degrees,
$\{\deg\mathcal{P}_n|n\in\mathbb{Z}_{\geq 0}\}\subsetneq
\mathbb{Z}_{\geq 0}$ \cite{gkm08,gkm08_2}.
By combining with (\romannumeral2), we can also consider the situation
in which second order differential equation is replaced with second order
difference equation.
These polynomials are called exceptional or multi-indexed orthogonal
polynomials and various examples have been obtained for classical orthogonal
polynomials
\cite{gkm08}--\cite{casoidrdqm}.
We distinguish the following two cases;
the set of missing degrees $\mathcal{I}=\mathbb{Z}_{\geq 0}\backslash
\{\text{deg}\,\mathcal{P}_n|n\in\mathbb{Z}_{\geq 0}\}$ is
case-(1): $\mathcal{I}=\{0,1,\ldots,\ell-1\}$, or
case-(2): $\mathcal{I}\neq\{0,1,\ldots,\ell-1\}$, where $\ell$ is a positive
integer. The situation of case-(1) is called stable in \cite{gkm11}.
In the case of finite systems such as ($q$-)Racah polynomials,
the index set $\mathbb{Z}_{\geq 0}$ is replaced by $\{0,1,\ldots,N\}$.
It is also possible to combine three directions (\romannumeral1),
(\romannumeral2) and (\romannumeral3), but such examples are not yet known.

Quantum mechanical formulation is useful for studying orthogonal polynomials.
We consider three kinds of quantum mechanical systems:
ordinary quantum mechanics (oQM), discrete quantum mechanics with pure
imaginary shifts (idQM) \cite{os13}--\cite{os24} and discrete quantum
mechanics with real shifts (rdQM) \cite{os12}--\cite{os34}.
Their features are the following:
\begin{center}
\begin{tabular}{c|ccl}
&Schr\"odinger eq.&variable $x$&examples of orthogonal polynomials\\[2pt]
\hline\\[-10pt]
oQM&differential eq.&continuous&Hermite, Laguerre, Jacobi\\
idQM&difference eq.&continuous&continuous Hahn, (Askey-)Wilson\\
rdQM&difference eq.&discrete&Hahn, ($q$-)Racah
\end{tabular}
\end{center}
In our previous works we have taken second order differential or difference
operators as Hamiltonians, but it is also allowed to take higher order
operators as Hamiltonians.
Exceptional and multi-indexed polynomials are obtained by applying the Darboux
transformations with appropriate seed solutions to the exactly solvable
quantum mechanical systems described by the classical orthogonal polynomials
in the Askey scheme.
When the virtual state wavefunctions are used as seed solutions,
the case-(1) multi-indexed polynomials are obtained \cite{os25,os27,os26}.
When the eigenstate and/or pseudo virtual state wavefunctions are used as
seed solutions, the case-(2) multi-indexed polynomials are obtained
\cite{os29}--\cite{casoidrdqm}.
Another method to obtain exceptional and multi-indexed polynomials is to use
the Krall-type polynomials
\cite{d13_2}--\cite{d15}.

In this paper we discuss dual polynomials of the case-(1) multi-indexed
($q$-)Racah polynomials.
Dual polynomials are introduced naturally for orthogonal polynomials of
a discrete variable \cite{ismail} and they are treated in the framework
of rdQM \cite{os12}--\cite{os34}.
The polynomial $\mathcal{P}_n(\eta(x))=\check{\mathcal{P}}_n(x)$ and its dual
polynomial $\mathcal{Q}_x(\mathcal{E}_n)=\check{\mathcal{Q}}_x(n)$ are
related as $\check{\mathcal{P}}_n(x)\propto\check{\mathcal{Q}}_x(n)$.
The roles of the variable and the label ($=$ degree for ordinary orthogonal
polynomials) are interchanged, and we have the following correspondence:
\begin{equation*}
  \begin{array}{rcl}
  \text{difference equation (w.r.t $x$) for $\check{\mathcal{P}}_n(x)$}
  &\leftrightarrow&
  \text{recurrence relation (w.r.t $x$) for $\check{\mathcal{Q}}_x(n)$},\\[2pt]
  \text{recurrence relation (w.r.t $n$) for $\check{\mathcal{P}}_n(x)$}
  &\leftrightarrow&
  \text{difference equation (w.r.t $n$) for $\check{\mathcal{Q}}_x(n)$}.
  \end{array}
\end{equation*}
The multi-indexed ($q$-)Racah polynomials satisfy the second order difference
equations \cite{os26}.
On the other hand, the multi-indexed polynomials do not satisfy the three
term recurrence relations, which characterize the ordinary orthogonal
polynomials \cite{ismail}, because they are not the ordinary orthogonal
polynomials.
They satisfy recurrence relations with more terms
\cite{stz10}--\cite{rrmiop4}, and such recurrence relations for the
multi-indexed ($q$-)Racah polynomials are studied recently \cite{rrmiop5}.
It is shown that the $M$-indexed ($q$-)Racah polynomials satisfy
various $1+2L$ ($L\geq M+1$) term recurrence relations with constant
coefficients.
Therefore dual polynomials of the multi-indexed ($q$-)Racah polynomials
satisfy the three term recurrence relations and various $2L$-th order
difference equations, namely they are ordinary orthogonal polynomials and
the Krall-type. The weight functions do not contain delta functions
(Kronecker deltas).
By using these dual multi-indexed ($q$-)Racah polynomials, we construct
new exactly solvable rdQM systems, whose Hamiltonians are not tridiagonal
but ``$(1+2L)$-diagonal''.
These quantum systems satisfy the closure relations \cite{os7,os12},
from which the creation and annihilation operators are obtained,
but they are not shape invariant.

This paper is organized as follows.
In section \ref{sec:miop} the essence of the multi-indexed ($q$-)Racah
polynomials are recapitulated.
In section \ref{sec:dmiopqR} we define the dual polynomials of the
multi-indexed ($q$-)Racah polynomials and present their properties.
In section \ref{sec:rdQM} we construct new exactly solvable rdQM systems
described by the dual multi-indexed ($q$-)Racah polynomials.
The closure relations and the creation and annihilation operators are
presented in \S\,\ref{sec:cr} and the shape invariance is discussed in
\S\,\ref{sec:si}. Some examples are given in \S\,\ref{sec:ex}.
Section \ref{sec:summary} is for a summary and comments.
In Appendix \ref{app:data} some basic data of the multi-indexed
($q$-)Racah polynomials are summarized for readers' convenience.

\section{Multi-indexed ($q$-)Racah Orthogonal Polynomials}
\label{sec:miop}

In this section we recapitulate the properties of the case-(1) multi-indexed
Racah (R) and $q$-Racah ($q$R) orthogonal polynomials \cite{os26,rrmiop5}.
We follow the notation of \cite{os26,rrmiop5}.
Various quantities depend on a set of parameters
$\bm{\lambda}=(\lambda_1,\lambda_2,\ldots)$ and their dependence is
expressed like, $f=f(\bm{\lambda})$, $f(x)=f(x;\bm{\lambda})$.
The parameter $q$ is $0<q<1$ and $q^{\bm{\lambda}}$ stands for
$q^{(\lambda_1,\lambda_2,\ldots)}=(q^{\lambda_1},q^{\lambda_2},\ldots)$.
See Appendix \ref{app:data} for the explicit forms of various quantities
($\check{\Xi}_{\mathcal{D}}(x)$, $\Xi_{\mathcal{D}}(\eta)$, 
$\check{P}_{\mathcal{D},n}(x)$, $P_{\mathcal{D},n}(\eta)$,
$\check{P}_n(x)$, $P_n(\eta)$,
$B(x)$, $D(x)$, $\tilde{\bm{\delta}}$, $\phi_0(x)$, 
$d_{\mathcal{D},n}$, $A_n$, $C_n$,
$\tilde{\mathcal{E}}_{\text{v}}$,
$I_{\bm{\lambda}}$).

The set of parameters $\bm{\lambda}=(\lambda_1,\lambda_2,\lambda_3,\lambda_4)$,
its shift $\bm{\delta}$ and $\kappa$ are
\begin{equation}
  \begin{array}{rl}
  \text{R}:&\bm{\lambda\,}=(a,b,c,d),\quad \bm{\delta}=(1,1,1,1),
  \quad\kappa=1,\\[5pt]
  \text{$q$R}:&q^{\bm{\lambda}}=(a,b,c,d),\quad \bm{\delta}=(1,1,1,1),
  \quad\kappa=q^{-1}.
  \end{array}
  \label{lambda}
\end{equation}
For $N\in\mathbb{Z}_{>0}$, we take $n_{\text{max}}=x_{\text{max}}=N$ and
\begin{equation}
  \text{R}:\ a=-N,\qquad
  \text{$q$R}:\ a=q^{-N},
  \label{a=-N}
\end{equation}
and assume the following parameter ranges:
\begin{equation}
  \begin{array}{rl}
  \text{R}:&0<d<a+b,\quad 0<c<1+d,\quad d+\max(\mathcal{D})+1<a+b\\[4pt]
  \text{$q$R}:&0<ab<d<1,\quad qd<c<1,\quad ab<dq^{\max(\mathcal{D})+1}.
  \end{array}
  \label{para}
\end{equation}
Here $\mathcal{D}=\{d_1,d_2,\ldots,d_M\}$
($d_1<d_2<\cdots<d_M$, $d_j\in\mathbb{Z}_{\geq 1}$) is the multi-index set.
The denominator polynomials $\Xi_{\mathcal{D}}(\eta)$ and the multi-indexed
($q$-)Racah polynomials $P_{\mathcal{D},n}(\eta)$
($n=0,1,\ldots,n_{\text{max}}$) are polynomials in the sinusoidal coordinate
$\eta$,
\begin{alignat}{2}
  \check{\Xi}_{\mathcal{D}}(x;\bm{\lambda})
  &\eqdef\Xi_{\mathcal{D}}\bigl(\eta(x;\bm{\lambda}+(M-1)\bm{\delta});
  \bm{\lambda}\bigr),&\quad\deg\Xi_{\mathcal{D}}(\eta)&=\ell_{\mathcal{D}},\\
  \check{P}_{\mathcal{D},n}(x;\bm{\lambda})
  &\eqdef P_{\mathcal{D},n}\bigl(\eta(x;\bm{\lambda}+M\bm{\delta});
  \bm{\lambda}\bigr),
  &\deg P_{\mathcal{D},n}(\eta)&=\ell_{\mathcal{D}}+n,
\end{alignat}
where $\ell_{\mathcal{D}}$ is
\begin{equation}
  \ell_{\mathcal{D}}\eqdef\sum_{j=1}^Md_j-\frac12M(M-1).
\end{equation}
The sinusoidal coordinates $\eta(x;\bm{\lambda})$ are
\begin{equation}
  \eta(x;\bm{\lambda})=\left\{
  \begin{array}{ll}
  x(x+d)&:\text{R}\\[2pt]
  (q^{-x}-1)(1-dq^x)&:\text{$q$R}
  \end{array}\right.,
  \label{eta}
\end{equation}
and the energy eigenvalues $\mathcal{E}_n(\bm{\lambda})$ are
\begin{equation}
  \mathcal{E}_n(\bm{\lambda})=\left\{
  \begin{array}{ll}
  n(n+\tilde{d})&:\text{R}\\
  (q^{-n}-1)(1-\tilde{d}q^n)&:\text{$q$R}
  \end{array}\right.,\quad
  \tilde{d}\eqdef\left\{
  \begin{array}{ll}
  a+b+c-d-1&:\text{R}\\[2pt]
  abcd^{-1}q^{-1}&:\text{$q$R}
  \end{array}\right..
  \label{En}
\end{equation}
The normalization of these quantities is
\begin{equation}
  \eta(0;\bm{\lambda})=\mathcal{E}_0(\bm{\lambda})=0,\quad
  \check{\Xi}_{\mathcal{D}}(0;\bm{\lambda})
  =\Xi_{\mathcal{D}}(0;\bm{\lambda})=1,\quad
  \check{P}_{\mathcal{D},n}(0;\bm{\lambda})
  =P_{\mathcal{D},n}(0;\bm{\lambda})=1.
\end{equation}
In the sequence $\check{P}_{\mathcal{D},n}(0;\bm{\lambda}),
\check{P}_{\mathcal{D},n}(1;\bm{\lambda}),
\ldots,\check{P}_{\mathcal{D},n}(x_{\text{max}};\bm{\lambda})$,
the sign changes $n$ times.
Note that
\begin{equation}
  \check{P}_{\mathcal{D},0}(x;\bm{\lambda})
  =\check{\Xi}_{\mathcal{D}}(x;\bm{\lambda}+\bm{\delta}),
  \label{PD0=XiD}
\end{equation}
and $\check{\Xi}_{\mathcal{D}}(x;\bm{\lambda})$ is positive for
$x=0,1,\ldots,x_{\text{max}}$.
We set
\begin{equation}
  P_{\mathcal{D},n}(\eta;\bm{\lambda})\eqdef0\ \ (n<0).
  \label{PDn<0=0}
\end{equation}
The original ($q$-)Racah polynomials $P_n(\eta)$ correspond to the ``$M=0$''
($\mathcal{D}=\emptyset$) case, $P_n(\eta)=P_{\emptyset,n}(\eta)$.
We remark that if we treat the parameter $a$ as an indeterminate,
$\check{P}_{\mathcal{D},n}(x)$ are defined for
$n\in\mathbb{Z}_{\geq 0}$ and $x\in\mathbb{C}$.
However, for the choice \eqref{a=-N} (we take the limit from an indeterminate
$a$ to $a$ in \eqref{a=-N}), $\check{P}_{\mathcal{D},n}(x)$ are well-defined
for $n\in\{0,1,\ldots,n_{\text{max}}\}$ and $x\in\mathbb{C}$, or
$n\in\mathbb{Z}_{>n_{\text{max}}}$ and $x\in\{0,1,\ldots,x_{\text{max}}\}$.

The Hamiltonian of the deformed system $\mathcal{H}_{\mathcal{D}}
=(\mathcal{H}_{\mathcal{D};x,y})_{0\leq x,y\leq x_{\text{max}}}$ is a
real symmetric matrix (a tridiagonal matrix in this case),
\begin{equation}
  \mathcal{H}_{\mathcal{D}}
  =-\sqrt{B_{\mathcal{D}}(x)}\,e^{\partial}\sqrt{D_{\mathcal{D}}(x)}
  -\sqrt{D_{\mathcal{D}}(x)}\,e^{-\partial}\sqrt{B_{\mathcal{D}}(x)}
  +B_{\mathcal{D}}(x)+D_{\mathcal{D}}(x),
  \label{HD}
\end{equation}
where matrices $e^{\pm\partial}$ are
$(e^{\pm\partial})_{x,y}=\delta_{x\pm 1,y}$ and the unit matrix
$\bm{1}=(\delta_{x,y})$ are suppressed.
The notation $f(x)Ag(x)$ (or $f(x)\circ A\circ g(x)$),
where $f(x)$ and $g(x)$ are functions of $x$ and
$A$ is a matrix $A=(A_{x,y})$, stands for a matrix whose $(x,y)$-element
is $f(x)A_{x,y}g(y)$.
Namely, it is a matrix product
$\text{diag}(f(0),f(1),\ldots f(x_{\text{max}}))\,A\,\text{diag}(g(0),g(1),
\ldots,g(x_{\text{max}}))$.
The notation $Af(x)$ stands for a vector whose $x$-th component is
$\sum\limits_{y=0}^{x_{\text{max}}}A_{x,y}f(y)$.
Note that the matrices $e^{\partial}$ and $e^{-\partial}$ are not inverse
to each other.
The potential functions $B_{\mathcal{D}}(x)$ and $D_{\mathcal{D}}(x)$ are
\begin{align}
  B_{\mathcal{D}}(x;\bm{\lambda})&=B(x;\bm{\lambda}+M\tilde{\bm{\delta}})\,
  \frac{\check{\Xi}_{\mathcal{D}}(x;\bm{\lambda})}
  {\check{\Xi}_{\mathcal{D}}(x+1;\bm{\lambda})}
  \frac{\check{\Xi}_{\mathcal{D}}(x+1;\bm{\lambda}+\bm{\delta})}
  {\check{\Xi}_{\mathcal{D}}(x;\bm{\lambda}+\bm{\delta})},\\
  D_{\mathcal{D}}(x;\bm{\lambda})&=D(x;\bm{\lambda}+M\tilde{\bm{\delta}})\,
  \frac{\check{\Xi}_{\mathcal{D}}(x+1;\bm{\lambda})}
  {\check{\Xi}_{\mathcal{D}}(x;\bm{\lambda})}
  \frac{\check{\Xi}_{\mathcal{D}}(x-1;\bm{\lambda}+\bm{\delta})}
  {\check{\Xi}_{\mathcal{D}}(x;\bm{\lambda}+\bm{\delta})},
\end{align}
which satisfy the boundary conditions
\begin{equation}
  D_{\mathcal{D}}(0;\bm{\lambda})=0,\quad
  B_{\mathcal{D}}(x_{\text{max}};\bm{\lambda})=0.
\end{equation}
The eigenvectors of the Hamiltonian are
\begin{align}
  &\phi_{\mathcal{D}\,n}(x;\bm{\lambda})=\psi_{\mathcal{D}}(x;\bm{\lambda})
  \check{P}_{\mathcal{D},n}(x;\bm{\lambda}),
  \ \ \psi_{\mathcal{D}}(x;\bm{\lambda})=
  \sqrt{\check{\Xi}_{\mathcal{D}}(1;\bm{\lambda})}\,
  \frac{\phi_0(x;\bm{\lambda}+M\tilde{\bm{\delta}})}
  {\sqrt{\check{\Xi}_{\mathcal{D}}(x;\bm{\lambda})\,
  \check{\Xi}_{\mathcal{D}}(x+1;\bm{\lambda})}},\\
  &\mathcal{H}_{\mathcal{D}}(\bm{\lambda})\phi_{\mathcal{D}\,n}(x;\bm{\lambda})
  =\mathcal{E}_n(\bm{\lambda})\phi_{\mathcal{D}\,n}(x;\bm{\lambda})
  \ \ (n=0,1,\ldots,n_{\text{max}}),
  \label{HDphiDn=}
\end{align}
and the normalization of $\psi_{\mathcal{D}}$ and $\phi_{\mathcal{D}\,n}$ is
$\psi_{\mathcal{D}}(0;\bm{\lambda})=\phi_{\mathcal{D}\,n}(0;\bm{\lambda})=1$.
Namely the multi-indexed ($q$-)Racah polynomials satisfy the second order
difference equations
\begin{equation}
  \widetilde{\mathcal{H}}_{\mathcal{D}}(\bm{\lambda})
  \check{P}_{\mathcal{D},n}(x;\bm{\lambda})
  =\mathcal{E}_n(\bm{\lambda})\check{P}_{\mathcal{D},n}(x;\bm{\lambda})
  \ \ (n=0,1,\ldots,n_{\text{max}}),
  \label{tHDcPDn=}
\end{equation}
where the similarity transformed Hamiltonian
$\widetilde{\mathcal{H}}_{\mathcal{D}}(\bm{\lambda})
=\psi_{\mathcal{D}}(x;\bm{\lambda})^{-1}\circ
\mathcal{H}_{\mathcal{D}}(\bm{\lambda})\circ
\psi_{\mathcal{D}}(x;\bm{\lambda})$ is
\begin{align}
  \widetilde{\mathcal{H}}_{\mathcal{D}}(\bm{\lambda})
  &=B(x;\bm{\lambda}+M\tilde{\bm{\delta}})\,
  \frac{\check{\Xi}_{\mathcal{D}}(x;\bm{\lambda})}
  {\check{\Xi}_{\mathcal{D}}(x+1;\bm{\lambda})}
  \biggl(\frac{\check{\Xi}_{\mathcal{D}}(x+1;\bm{\lambda}+\bm{\delta})}
  {\check{\Xi}_{\mathcal{D}}(x;\bm{\lambda}+\bm{\delta})}-e^{\partial}
  \biggr)\n
  &\quad+D(x;\bm{\lambda}+M\tilde{\bm{\delta}})\,
  \frac{\check{\Xi}_{\mathcal{D}}(x+1;\bm{\lambda})}
  {\check{\Xi}_{\mathcal{D}}(x;\bm{\lambda})}
  \biggl(\frac{\check{\Xi}_{\mathcal{D}}(x-1;\bm{\lambda}+\bm{\delta})}
  {\check{\Xi}_{\mathcal{D}}(x;\bm{\lambda}+\bm{\delta})}-e^{-\partial}
  \biggr).
\end{align}

The orthogonality relations of the multi-indexed ($q$-)Racah polynomials are
\begin{equation}
  \sum_{x=0}^{x_{\text{max}}}
  \frac{\psi_{\mathcal{D}}(x;\bm{\lambda})^2}
  {\check{\Xi}_{\mathcal{D}}(1;\bm{\lambda})}
  \check{P}_{\mathcal{D},n}(x;\bm{\lambda})
  \check{P}_{\mathcal{D},m}(x;\bm{\lambda})
  =\frac{\delta_{nm}}{d_{\mathcal{D},n}(\bm{\lambda})^2}
  \ \ (n,m=0,1,\ldots,n_{\text{max}}).
  \label{orthomiop}
\end{equation}
We remark that
\begin{equation}
  \frac{d_{\mathcal{D},n}(\bm{\lambda})}{d_{\mathcal{D},0}(\bm{\lambda})}
  =\biggl(\prod_{m=0}^{n-1}\frac{A_m(\bm{\lambda})}{C_{m+1}(\bm{\lambda})}
  \cdot\prod_{j=1}^M
  \frac{\mathcal{E}_n(\bm{\lambda})-\tilde{\mathcal{E}}_{d_j}(\bm{\lambda})}
  {-\tilde{\mathcal{E}}_{d_j}(\bm{\lambda})}
  \biggr)^{\frac12},
  \label{dDn/dD0}
\end{equation}
where $A_n$ and $C_n$ are the coefficients of the three term recurrence
relations for the original ($q$-)Racah polynomials $P_n(\eta)$ and
$\tilde{\mathcal{E}}_{\text{v}}$ is the virtual state energy.

The multi-indexed ($q$-)Racah polynomials satisfy the recurrence relations
with constant coefficients \cite{rrmiop5}.
We have the following results:
\begin{thm}$\!${\rm\cite{rrmiop5}}
\label{thm:rr}
For any polynomial $Y(\eta)(\neq 0)$, we take
$X(\eta)=X(\eta;\bm{\lambda})=X^{\mathcal{D},Y}(\eta;\bm{\lambda})$ as
\begin{equation}
  X(\eta)=I_{\bm{\lambda}+M\bm{\delta}}
  \bigl[\Xi_{\mathcal{D}}Y\bigr](\eta),\quad
  \deg X(\eta)=L=\ell_{\mathcal{D}}+\deg Y(\eta)+1,
  \label{X=I[XiY]}
\end{equation}
where $\Xi_{\mathcal{D}}Y$ means a polynomial
$(\Xi_{\mathcal{D}}Y)(\eta)=\Xi_{\mathcal{D}}(\eta)Y(\eta)$,
and define $\check{X}(x)=\check{X}(x;\bm{\lambda})$ by
\begin{align}
  \check{X}(x;\bm{\lambda})&\eqdefrm
  X\bigl(\eta(x;\bm{\lambda}+M\bm{\delta});\bm{\lambda}\bigr)
  \ \ (x\in\mathbb{C})
  \label{cX}\\
  &=\sum_{j=1}^x\bigl(\eta(j;\bm{\lambda}+M\bm{\delta})
  -\eta(j-1;\bm{\lambda}+M\bm{\delta})\bigr)\n
  &\qquad\quad\times\check{\Xi}_{\mathcal{D}}(j;\bm{\lambda})
  Y\bigl(\eta(j;\bm{\lambda}+(M-1)\bm{\delta})\bigr)
  \ \ (x\in\mathbb{Z}_{\geq 0}).
  \label{cX2}
\end{align}
Then the multi-indexed ($q$-)Racah polynomials $P_{\mathcal{D},n}(\eta)$
satisfy $1+2L$ term recurrence relations with constant coefficients:
\begin{equation}
  \check{X}(x;\bm{\lambda})\check{P}_{\mathcal{D},n}(x;\bm{\lambda})
  =\sum_{k=-\min(L,n)}^{\min(L,N-n)}
  r_{n,k}^{X,\mathcal{D}}(\bm{\lambda})
  \check{P}_{\mathcal{D},n+k}(x;\bm{\lambda})
  \ \ \Bigl(\begin{array}{ll}
  n=0,1,\ldots,n_{\text{\rm max}}\\
  x=0,1,\ldots,x_{\text{\rm max}}
  \end{array}\Bigr).
  \label{cXcP}
\end{equation}
\end{thm}
{\bf Remark}\,
For $L>\frac12N$, number of terms is not $1+2L$ but $N+1$.
Unless $N-L+1\leq n\leq N$, \eqref{cXcP} is an equation as a polynomial,
namely it holds for $x\in\mathbb{C}$.
On the other hand, for $N-L+1\leq n\leq N$, \eqref{cXcP} holds only for
$x=0,1,\ldots,x_{\text{max}}$.

We note that
the overall normalization and the constant term of $X(\eta)$ are not important,
because the change of the former induces that of the overall normalization
of $r_{n,k}^{X,\mathcal{D}}$ and the shift of the latter induces that of
$r_{n,0}^{X,\mathcal{D}}$.
The constant term of $X(\eta)$ is chosen as $X(0)=0$.
There are the following relations among the coefficients
$r_{n,k}^{X,\mathcal{D}}$ \cite{rrmiop5}
\begin{equation}
  r_{n+k,-k}^{X,\mathcal{D}}(\bm{\lambda})
  =\frac{d_{\mathcal{D},n}(\bm{\lambda})^2}
  {d_{\mathcal{D},n+k}(\bm{\lambda})^2}\,
  r_{n,k}^{X,\mathcal{D}}(\bm{\lambda})
  \ \ (1\leq k\leq L;n+k\leq n_{\text{max}}).
  \label{r-k=rk}
\end{equation}
Direct verification of this theorem is rather straightforward for lower
$M$ and smaller $d_j$, $n$, $\deg Y$ and $N$, by a computer algebra system,
e.g.\! Mathematica.
The coefficients $r_{n,k}^{X,\mathcal{D}}$ are explicitly obtained for
small $d_j$ and $n$. However, to obtain the closed expression of
$r_{n,k}^{X,\mathcal{D}}$ for general $n$ is not an easy task even for
small $d_j$, and it is a different kind of problem.
Since $Y(\eta)$ is arbitrary, we obtain infinitely many recurrence relations.
Although not all of them are independent, the relations among them are unclear.
Note that $L\geq M+1$ because of $\ell_{\mathcal{D}}\geq M$.
The minimal degree one, which corresponds to $Y(\eta)=1$, is
$X_{\text{min}}(\eta)=I_{\bm{\lambda}+M\bm{\delta}}
\bigl[\Xi_{\mathcal{D}}\bigr](\eta)$,
$\deg X_{\text{min}}(\eta)=\ell_{\mathcal{D}}+1$.

\section{Dual Polynomials of the Multi-Indexed ($q$-)Racah Polynomials}
\label{sec:dmiopqR}

For ordinary orthogonal polynomials, a discrete orthogonality relation of
a system of polynomials induces an orthogonality relation for the dual system
where the role of the variable and the degree are interchanged \cite{ismail}
(see also \cite{os12}--\cite{os34}).
In this section we consider dual polynomials of the multi-indexed ($q$-)Racah
polynomials, where the degree is replaced by the number of sign changes.

Corresponding to the multi-indexed ($q$-)Racah polynomials
$\check{P}_{\mathcal{D},n}(x;\bm{\lambda})$,
let us define $\check{Q}_{\mathcal{D},x}(n;\bm{\lambda})$ by
\begin{equation}
  \check{Q}_{\mathcal{D},x}(n;\bm{\lambda})
  \eqdef\frac{\check{P}_{\mathcal{D},n}(x;\bm{\lambda})}
  {\check{P}_{\mathcal{D},0}(x;\bm{\lambda})}
  \ \ \Bigl(\begin{array}{ll}
  n=0,1,\ldots,n_{\text{max}}\\
  x=0,1,\ldots,x_{\text{max}}
  \end{array}\Bigr).
  \label{QDxn}
\end{equation}
(Remark that if the parameter $a$ is treated as an indeterminate,
$\check{Q}_{\mathcal{D},x}(n;\bm{\lambda})$ is defined for
$n,x\in\mathbb{Z}_{\geq 0}$.)
Then we have
\begin{equation}
  \check{Q}_{\mathcal{D},x}(0;\bm{\lambda})=1,\quad
  \check{Q}_{\mathcal{D},0}(n;\bm{\lambda})=1.
\end{equation}
Orthogonality relations \eqref{orthomiop} are rewritten as
\begin{equation}
  \sum_{x=0}^{x_{\text{max}}}\hat{\phi}_{\mathcal{D}\,n}(x;\bm{\lambda})
  \hat{\phi}_{\mathcal{D}\,m}(x;\bm{\lambda})=\delta_{nm}
  \ \ (n,m=0,1,\ldots,n_{\text{max}}),
  \label{orthohatphi}
\end{equation}
where the normalized eigenvectors $\hat{\phi}_{\mathcal{D}\,n}(x;\bm{\lambda})$
are
\begin{equation}
  \hat{\phi}_{\mathcal{D}\,n}(x;\bm{\lambda})\eqdef
  \frac{d_{\mathcal{D},n}(\bm{\lambda})}
  {\sqrt{\check{\Xi}_{\mathcal{D}}(1;\bm{\lambda})}}
  \psi_{\mathcal{D}}(x;\bm{\lambda})\check{P}_{\mathcal{D},n}(x;\bm{\lambda})
  =\frac{\phi_{\mathcal{D}\,0}(x;\bm{\lambda})}
  {\sqrt{\check{\Xi}_{\mathcal{D}}(1;\bm{\lambda})}}
  d_{\mathcal{D},n}(\bm{\lambda})\check{Q}_{\mathcal{D},x}(n;\bm{\lambda}).
\end{equation}
Since the matrix size is finite $n_{\text{max}}=x_{\text{max}}=N$,
\eqref{orthohatphi} implies
\begin{equation}
  \sum_{n=0}^{n_{\text{max}}}\hat{\phi}_{\mathcal{D}\,n}(x;\bm{\lambda})
  \hat{\phi}_{\mathcal{D}\,n}(y;\bm{\lambda})=\delta_{xy}
  \ \ (x,y=0,1,\ldots,x_{\text{max}}),
  \label{orthohatphi2}
\end{equation}
namely dual orthogonality relations
\begin{equation}
  \sum_{n=0}^{n_{\text{max}}}
  \frac{d_{\mathcal{D},n}(\bm{\lambda})^2}
  {\check{\Xi}_{\mathcal{D}}(1;\bm{\lambda})}
  \check{Q}_{\mathcal{D},x}(n;\bm{\lambda})
  \check{Q}_{\mathcal{D},y}(n;\bm{\lambda})
  =\frac{\delta_{xy}}{\phi_{\mathcal{D}\,0}(x;\bm{\lambda})^2}
  \ \ (x,y=0,1,\ldots,x_{\text{max}}).
  \label{orthomiop2}
\end{equation}

The second order difference equations for
$\check{P}_{\mathcal{D},n}(x;\bm{\lambda})$ \eqref{tHDcPDn=} are rewritten
as the three term recurrence relations
for $\check{Q}_{\mathcal{D},x}(n;\bm{\lambda})$,
\begin{align}
  \mathcal{E}_n(\bm{\lambda})\check{Q}_{\mathcal{D},x}(n;\bm{\lambda})
  &=A^{\text{dual}}_{\mathcal{D},x}(\bm{\lambda})
  \check{Q}_{\mathcal{D},x+1}(n;\bm{\lambda})
  +B^{\text{dual}}_{\mathcal{D},x}(\bm{\lambda})
  \check{Q}_{\mathcal{D},x}(n;\bm{\lambda})
  +C^{\text{dual}}_{\mathcal{D},x}(\bm{\lambda})
  \check{Q}_{\mathcal{D},x-1}(n;\bm{\lambda})\n
  &\qquad\qquad\qquad\qquad\qquad
  (n=0,1,\ldots,n_{\text{max}}\,;\,x=0,1,\ldots,x_{\text{max}}),
  \label{3termcQ}
\end{align}
where we have used \eqref{PD0=XiD}.
Here $A^{\text{dual}}_{\mathcal{D},x}$, $B^{\text{dual}}_{\mathcal{D},x}$
and $C^{\text{dual}}_{\mathcal{D},x}$ are
\begin{equation}
  A^{\text{dual}}_{\mathcal{D},x}(\bm{\lambda})
  \eqdef-B_{\mathcal{D}}(x;\bm{\lambda}),
  \ \ C^{\text{dual}}_{\mathcal{D},x}(\bm{\lambda})
  \eqdef-D_{\mathcal{D}}(x;\bm{\lambda}),
  \ \ B^{\text{dual}}_{\mathcal{D},x}(\bm{\lambda})
  \eqdef-A^{\text{dual}}_{\mathcal{D},x}(\bm{\lambda})
  -C^{\text{dual}}_{\mathcal{D},x}(\bm{\lambda}),
\end{equation}
and satisfy the boundary conditions
\begin{equation}
  C^{\text{dual}}_{\mathcal{D},0}(\bm{\lambda})=0,\quad
  A^{\text{dual}}_{\mathcal{D},x_{\text{max}}}(\bm{\lambda})=0.
\end{equation}
This means that $\check{Q}_{\mathcal{D},x}(n;\bm{\lambda})$ are generated
by the three term recurrence relations \eqref{3termcQ} with the initial
conditions
\begin{equation}
  \check{Q}_{\mathcal{D},0}(n;\bm{\lambda})=1,\quad
  \check{Q}_{\mathcal{D},-1}(n;\bm{\lambda})\eqdef 0.
  \label{initialQ}
\end{equation}
Therefore $\check{Q}_{\mathcal{D},x}(n;\bm{\lambda})$ are polynomials in
$\mathcal{E}_n(\bm{\lambda})$,
\begin{equation}
  \check{Q}_{\mathcal{D},x}(n;\bm{\lambda})
  \eqdef Q_{\mathcal{D},x}\bigl(\mathcal{E}_n(\bm{\lambda});\bm{\lambda}\bigr),
  \quad\deg Q_{\mathcal{D},x}(\mathcal{E})=x.
\end{equation}
These polynomials $Q_{\mathcal{D},x}(\mathcal{E})$ are ordinary orthogonal
polynomials.
So, the sign changes $x$ times in the sequence
$\check{Q}_{\mathcal{D},x}(0;\bm{\lambda}),
\check{Q}_{\mathcal{D},x}(1;\bm{\lambda}),
\ldots,\check{Q}_{\mathcal{D},x}(n_{\text{max}};\bm{\lambda})$.
We call $Q_{\mathcal{D},x}(\mathcal{E})$ as the dual multi-indexed
($q$-)Racah polynomials.
Correspondence of this duality is
\begin{equation}
  x\leftrightarrow n,\quad
  \eta(x)\leftrightarrow\mathcal{E}_n,\quad
  P_{\mathcal{D},n}(\eta)\leftrightarrow Q_{\mathcal{D},x}(\mathcal{E}),\quad
  \frac{\phi_{\mathcal{D}\,0}(x)}{\phi_{\mathcal{D}\,0}(0)}\leftrightarrow
  \frac{d_{\mathcal{D},n}}{d_{\mathcal{D},0}}.
\end{equation}
The multi-indexed ($q$-)Racah polynomials $P_{\mathcal{D},n}(\eta)$ and
their dual polynomials $Q_{\mathcal{D},x}(\mathcal{E})$ are different
polynomials.
This contrasts with the original ($q$-)Racah cases.
The ($q$-)Racah polynomials and their dual polynomials are same
polynomials with the parameter correspondence
$(a,b,c,d)\leftrightarrow(a,b,c,\tilde{d})$.
We remark that if we treat the parameter $a$ as an indeterminate,
the dual multi-indexed ($q$-)Racah polynomials
$Q_{\mathcal{D},x}(\mathcal{E})$ are defined for $x\in\mathbb{Z}_{\ge 0}$
and $\mathcal{E}\in\mathbb{C}$ by the three term recurrence relations
\begin{equation}
  \mathcal{E}Q_{\mathcal{D},x}(\mathcal{E};\bm{\lambda})
  =A^{\text{dual}}_{\mathcal{D},x}(\bm{\lambda})
  Q_{\mathcal{D},x+1}(\mathcal{E};\bm{\lambda})
  +B^{\text{dual}}_{\mathcal{D},x}(\bm{\lambda})
  Q_{\mathcal{D},x}(\mathcal{E};\bm{\lambda})
  +C^{\text{dual}}_{\mathcal{D},x}(\bm{\lambda})
  Q_{\mathcal{D},x-1}(\mathcal{E};\bm{\lambda}),
  \label{3termQ}
\end{equation}
with the initial condition $Q_{\mathcal{D},0}(\mathcal{E})=1$ and
$Q_{\mathcal{D},-1}(\mathcal{E})\eqdef 0$.
For \eqref{a=-N}, we have $A^{\text{dual}}_{\mathcal{D},N}(\bm{\lambda})=0$.

The $1+2L$ term recurrence relations with constant coefficients for
$\check{P}_{\mathcal{D},n}(x;\bm{\lambda})$ \eqref{cXcP} are rewritten as
the $2L$-th order difference equations for
$\check{Q}_{\mathcal{D},x}(n;\bm{\lambda})$,
\begin{equation}
  \sum_{k=-\min(L,n)}^{\min(L,N-n)}r_{n,k}^{X,\mathcal{D}}(\bm{\lambda})
  \check{Q}_{\mathcal{D},x}(n+k;\bm{\lambda})
  =\check{X}(x;\bm{\lambda}\bigr)
  \check{Q}_{\mathcal{D},x}(n;\bm{\lambda})
  \ \ \Bigl(\begin{array}{ll}
  n=0,1,\ldots,n_{\text{max}}\\
  x=0,1,\ldots,x_{\text{max}}
  \end{array}\Bigr).
  \label{XQ}
\end{equation}
(Remark: For $L>\frac12N$, the order is not $2L$ but $N$.)
Therefore the dual multi-indexed ($q$-)Racah polynomials
$Q_{\mathcal{D},x}(\mathcal{E})$ are the Krall-type polynomials.
Since we can take various $X=X^{\mathcal{D},Y}$ for a multi-indexed set
$\mathcal{D}$, these Krall-type polynomials $Q_{\mathcal{D},x}(\mathcal{E})$
satisfy various difference equations of order $2L\geq 2M+2$.

In the $q\to1$ limit, the $q$-Racah polynomial reduced to the Racah polynomial
\cite{kls}: $\lim\limits_{q\to1}\check{P}^{\text{$q$R}}_n(x;\bm{\lambda})
=\check{P}^{\text{R}}_n(x;\bm{\lambda})$.
Similarly, the (dual) multi-indexed $q$-Racah polynomials reduce to the (dual)
multi-indexed Racah polynomials: 
$\lim\limits_{q\to1}\check{P}^{\text{$q$R}}_{\mathcal{D},n}(x;\bm{\lambda})
=\check{P}^{\text{R}}_{\mathcal{D},n}(x;\bm{\lambda})$ and
$\lim\limits_{q\to1}\check{Q}^{\text{$q$R}}_{\mathcal{D},x}(n;\bm{\lambda})
=\check{Q}^{\text{R}}_{\mathcal{D},x}(n;\bm{\lambda})$.

\section{New Exactly Solvable rdQM Systems}
\label{sec:rdQM}

In this section we present new exactly solvable rdQM systems, whose
eigenvectors are described by the dual multi-indexed ($q$-)Racah polynomials.
Unlike the previous section, we assume that the coordinate is $x$ and
the label of states is $n$ as usual.
Although tridiagonal matrices have been considered in our previous papers
\cite{os12,os22,os34,casoidrdqm}, the Hamiltonian of rdQM is not restricted
to tridiagonal matrices and any real symmetric matrix is allowed.

Let us fix the multi-index set $\mathcal{D}=\{d_1,d_2,\ldots,d_M\}$
($d_1<d_2<\cdots<d_M$, $d_j\in\mathbb{Z}_{\geq 1}$).
We take a polynomial $X(\eta)=X^{\mathcal{D},Y}(\eta)$ \eqref{X=I[XiY]} and
assume that $Y(\eta)(\neq 0)$ is a polynomial with real non-negative
coefficients.
For each $X(\eta)$, we define the Hamiltonian
$\mathcal{H}^{X\,\text{dual}}_{\mathcal{D}}(\bm{\lambda})=
\bigl(\mathcal{H}^{X\,\text{dual}}_{\mathcal{D};x,y}(\bm{\lambda})
\bigr)_{0\leq x,y\leq x_{\text{max}}}$ by
\begin{equation}
  \mathcal{H}^{X\,\text{dual}}_{\mathcal{D}}(\bm{\lambda})\eqdef
  \sum_{k=-\min(L,x)}^{\min(L,N-x)}
  r_{x,k}^{X,\mathcal{D}}(\bm{\lambda})
  \frac{d_{\mathcal{D},x}(\bm{\lambda})}{d_{\mathcal{D},x+k}(\bm{\lambda})}
  e^{k\partial},
  \label{HDXdual}
\end{equation}
where matrices $e^{k\partial}$ ($k\in\mathbb{Z}$) are defined by
\begin{equation}
  e^{k\partial}\eqdef\left\{
  \begin{array}{ll}
  (e^{\partial})^k&(k\geq 0)\\
  (e^{-\partial})^{-k}&(k<0)
  \end{array}\right.,
  \ \ \text{namely}\ \ (e^{k\partial})_{x,y}=\delta_{x+k,y}.
\end{equation}
This Hamiltonian $\mathcal{H}^{X\,\text{dual}}_{\mathcal{D}}$ is a band
matrix with lower and upper bandwidth $L$, namely a ``$(1+2L)$-diagonal''
matrix, e.g.\,$L=1$ ($\Leftrightarrow M=0,\mathcal{D}=\emptyset$):
tridiagonal matrix, $L=2$: pentadiagonal matrix, etc.
(For $L>N$, it is ``$(1+2N)$-diagonal''.)
By using \eqref{dDn/dD0} (with $n\rightarrow x$), explicit forms of
$d_{\mathcal{D},x}/d_{\mathcal{D},x+k}$ with $0\leq x+k\leq N$ are
(convention: $\prod_{i=n}^{n-1}*=1$)
\begin{equation}
  \frac{d_{\mathcal{D},x}(\bm{\lambda})}{d_{\mathcal{D},x+k}(\bm{\lambda})}
  =\biggl(\prod_{j=1}^M
  \frac{\mathcal{E}_x(\bm{\lambda})-\tilde{\mathcal{E}}_{d_j}(\bm{\lambda})}
  {\mathcal{E}_{x+k}(\bm{\lambda})-\tilde{\mathcal{E}}_{d_j}(\bm{\lambda})}
  \biggr)^{\frac12}
  \times\left\{
  \begin{array}{ll}
  {\displaystyle
  \biggl(\prod_{i=1}^k\frac{C_{x+i}(\bm{\lambda})}{A_{x+i-1}(\bm{\lambda})}
  \biggr)^{\frac12}}
  &(0\leq k\leq L)\\[12pt]
  {\displaystyle
  \biggl(\prod_{i=1}^{-k}\frac{A_{x-i}(\bm{\lambda})}{C_{x+1-i}(\bm{\lambda})}
  \biggr)^{\frac12}}
  &(-L\leq k\leq-1)
  \end{array}\right..
\end{equation}
This Hamiltonian $\mathcal{H}^{X\,\text{dual}}_{\mathcal{D}}$ is real
symmetric because of \eqref{r-k=rk} (with $n\rightarrow x$).

Multiplying \eqref{XQ} (with replacement $x\leftrightarrow n$) by
$d_{\mathcal{D},x}(\bm{\lambda})/d_{\mathcal{D},0}(\bm{\lambda})$,
we have
\begin{align}
  &\sum_{k=-\min(L,x)}^{\min(L,N-x)}r_{x,k}^{X,\mathcal{D}}(\bm{\lambda})
  \frac{d_{\mathcal{D},x}(\bm{\lambda})}{d_{\mathcal{D},x+k}(\bm{\lambda})}
  \frac{d_{\mathcal{D},x+k}(\bm{\lambda})}{d_{\mathcal{D},0}(\bm{\lambda})}
  \check{Q}_{\mathcal{D},n}(x+k;\bm{\lambda})
  =\check{X}(n;\bm{\lambda}\bigr)
  \frac{d_{\mathcal{D},x}(\bm{\lambda})}{d_{\mathcal{D},0}(\bm{\lambda})}
  \check{Q}_{\mathcal{D},n}(x;\bm{\lambda})\n
  &\qquad\qquad\qquad\qquad\qquad\qquad\qquad\qquad
  (n=0,1,\ldots,n_{\text{max}}\,;\,x=0,1,\ldots,x_{\text{max}}).
\end{align}
Therefore eigenvectors of the Hamiltonian
$\mathcal{H}^{X\,\text{dual}}_{\mathcal{D}}(\bm{\lambda})$,
\begin{equation}
  \mathcal{H}^{X\,\text{dual}}_{\mathcal{D}}(\bm{\lambda})
  \phi^{\text{dual}}_{\mathcal{D}\,n}(x;\bm{\lambda})
  =\mathcal{E}^{X\,\text{dual}}_{\mathcal{D},n}(\bm{\lambda})
  \phi^{\text{dual}}_{\mathcal{D}\,n}(x;\bm{\lambda})
  \ \ (n=0,1,\ldots,n_{\text{max}}),
  \label{dualHphiDn=}
\end{equation}
are given by
\begin{align}
  \phi^{\text{dual}}_{\mathcal{D}\,n}(x;\bm{\lambda})
  &\eqdef\frac{d_{\mathcal{D},x}(\bm{\lambda})}{d_{\mathcal{D},0}
  (\bm{\lambda})}\check{Q}_{\mathcal{D},n}(x;\bm{\lambda})
  \ \ \Bigl(\begin{array}{ll}
  n=0,1,\ldots,n_{\text{max}}\\
  x=0,1,\ldots,x_{\text{max}}
  \end{array}\Bigr),\\
  \mathcal{E}^{X\,\text{dual}}_{\mathcal{D},n}(\bm{\lambda})
  &\eqdef\check{X}(n;\bm{\lambda}\bigr),\qquad
  \check{Q}_{\mathcal{D},n}(x;\bm{\lambda})
  =Q_{\mathcal{D},n}\bigl(\mathcal{E}_x(\bm{\lambda});\bm{\lambda}\bigr).
  \label{EXdualDn}
\end{align}
Their orthogonality relations are obtained from \eqref{orthomiop2}:
\begin{equation}
  \sum_{x=0}^{x_{\text{max}}}
  \phi^{\text{dual}}_{\mathcal{D}\,n}(x;\bm{\lambda})
  \phi^{\text{dual}}_{\mathcal{D}\,m}(x;\bm{\lambda})
  =\frac{\check{\Xi}_{\mathcal{D}}(1;\bm{\lambda})\delta_{nm}}
  {d_{\mathcal{D},0}(\bm{\lambda})^2\,
  \phi_{\mathcal{D}\,0}(n;\bm{\lambda})^2}
  \ \ (n,m=0,1,\ldots,n_{\text{max}}).
\end{equation}
Since each term of the sum in \eqref{cX2} is positive for 
$x=1,2,\ldots,x_{\text{max}}$, we have
\begin{equation}
  0=\check{X}(0;\bm{\lambda})<\check{X}(1;\bm{\lambda})<\cdots<
  \check{X}(x_{\text{\rm max}};\bm{\lambda}).
  \label{Xincrease}
\end{equation}
Therefore the energy eigenvalues
$\mathcal{E}^{X\,\text{dual}}_{\mathcal{D},n}(\bm{\lambda})$ satisfy
\begin{equation}
  0=\mathcal{E}^{X\,\text{\rm dual}}_{\mathcal{D},0}(\bm{\lambda})<
  \mathcal{E}^{X\,\text{\rm dual}}_{\mathcal{D},1}(\bm{\lambda})<\cdots<
  \mathcal{E}^{X\,\text{\rm dual}}_{\mathcal{D},n_{\text{\rm max}}}
  (\bm{\lambda}),
  \label{Eincrease}
\end{equation}
and the Hamiltonian $\mathcal{H}^{X\,\text{dual}}_{\mathcal{D}}$ is
positive semi-definite.

For a multi-index set $\mathcal{D}$, we can take various $X=X^{\mathcal{D},Y}$,
because a polynomial $Y$ with real non-negative coefficients is arbitrary.
The Hamiltonian $\mathcal{H}^{X\,\text{dual}}_{\mathcal{D}}$ and energy
eigenvalues $\mathcal{E}^{X\,\text{dual}}_{\mathcal{D},n}$ depend on $X$,
but the eigenvectors $\phi^{\text{dual}}_{\mathcal{D}\,n}(x)$ do not.
Hence $\phi^{\text{dual}}_{\mathcal{D}\,n}(x)$ are simultaneous eigenvectors
of various $\mathcal{H}^{X\,\text{dual}}_{\mathcal{D}}$.
In other words, the Hamiltonians associated with $X_1=X^{\mathcal{D},Y_1}$
and $X_2=X^{\mathcal{D},Y_2}$ commute with each other,
\begin{equation}
  \bigl[\mathcal{H}^{X_1\,\text{dual}}_{\mathcal{D}},
  \mathcal{H}^{X_2\,\text{dual}}_{\mathcal{D}}\bigr]=0.
  \label{[HX1,HX2]=0}
\end{equation}

By the similarity transformation in terms of the ground state eigenvector
$\phi^{\text{dual}}_{\mathcal{D}\,0}(x)$, the Schr\"odinger equation
\eqref{dualHphiDn=} is rewritten as
\begin{align}
  &\widetilde{\mathcal{H}}^{X\,\text{dual}}_{\mathcal{D}}(\bm{\lambda})
  \check{Q}_{\mathcal{D},n}(x;\bm{\lambda})
  =\mathcal{E}^{X\,\text{dual}}_{\mathcal{D},n}(\bm{\lambda})
  \check{Q}_{\mathcal{D},n}(x;\bm{\lambda})
  \ \ (n=0,1,\ldots,n_{\text{max}}),
  \label{dualtHQDn=}\\
  &\widetilde{\mathcal{H}}^{X\,\text{dual}}_{\mathcal{D}}(\bm{\lambda})
  =\phi^{\text{dual}}_{\mathcal{D}\,0}(x;\bm{\lambda})^{-1}\circ
  \mathcal{H}^{X\,\text{dual}}_{\mathcal{D}}(\bm{\lambda})\circ
  \phi^{\text{dual}}_{\mathcal{D}\,0}(x;\bm{\lambda})
  =\sum_{k=-\min(L,x)}^{\min(L,N-x)}
  r_{x,k}^{X,\mathcal{D}}(\bm{\lambda})e^{k\partial}.
  \label{dualtH}
\end{align}

Multiplying \eqref{3termcQ} ($x\leftrightarrow n$) by
$d_{\mathcal{D},x}(\bm{\lambda})/d_{\mathcal{D},0}(\bm{\lambda})$,
we obtain the three term recurrence relations for the eigenvectors
$\phi^{\text{dual}}_{\mathcal{D}\,n}(x;\bm{\lambda})$,
\begin{align}
  &\mathcal{E}_x(\bm{\lambda})
  \phi^{\text{dual}}_{\mathcal{D}\,n}(x;\bm{\lambda})
  =A^{\text{dual}}_{\mathcal{D},n}(\bm{\lambda})
  \phi^{\text{dual}}_{\mathcal{D}\,n+1}(x;\bm{\lambda})
  +B^{\text{dual}}_{\mathcal{D},n}(\bm{\lambda})
  \phi^{\text{dual}}_{\mathcal{D}\,n}(x;\bm{\lambda})
  +C^{\text{dual}}_{\mathcal{D},n}(\bm{\lambda})
  \phi^{\text{dual}}_{\mathcal{D}\,n-1}(x;\bm{\lambda})\n
  &\qquad\qquad\qquad\qquad\qquad\qquad\qquad\qquad
  (n=0,1,\ldots,n_{\text{max}}\,;\,x=0,1,\ldots,x_{\text{max}}).
  \label{dualphiDn3term}
\end{align}

\subsection{Closure relation}
\label{sec:cr}

Corresponding to the three term recurrence relations \eqref{dualphiDn3term},
the Hamiltonian $\mathcal{H}^{X\,\text{dual}}_{\mathcal{D}}$
and the sinusoidal coordinate $\mathcal{E}_x$ are expected to satisfy the
ordinary closure relation \cite{os7,os12}
(closure relation of order 2 \cite{rrmiop4}),
\begin{equation}
  \bigl[\mathcal{H}^{X\,\text{dual}}_{\mathcal{D}},
  [\mathcal{H}^{X\,\text{dual}}_{\mathcal{D}},\underline{\mathcal{E}}\,]\bigr]
  =\underline{\mathcal{E}}\,R_0(\mathcal{H}^{X\,\text{dual}}_{\mathcal{D}})
  +[\mathcal{H}^{X\,\text{dual}}_{\mathcal{D}},\underline{\mathcal{E}}\,]
  R_1(\mathcal{H}^{X\,\text{dual}}_{\mathcal{D}})
  +R_{-1}(\mathcal{H}^{X\,\text{dual}}_{\mathcal{D}}),
  \label{cr}
\end{equation}
where $\underline{\mathcal{E}}$ is a diagonal matrix
$\underline{\mathcal{E}}
=(\mathcal{E}_x\delta_{x,y})_{0\leq x,y\leq x_{\text{max}}}$
and $R_i(z)$'s are polynomials in $z$.
(In the notation used in \eqref{HD}, this matrix $\underline{\mathcal{E}}$
is expressed as $\mathcal{E}_x{\bf 1}$ or simply $\mathcal{E}_x$.)
For the original ($q$-)Racah systems, the degrees of $R_i(z)$ are
$(\deg R_0,\deg R_1,\deg R_{-1})=(2,1,2)$.
For the dual multi-indexed ($q$-)Racah systems, the degrees of $R_i(x)$ need
to be much higher.
This is because the expression of
$\mathcal{E}^{X\,\text{dual}}_{\mathcal{D},n}$ is more complicated than
that of $\mathcal{E}_n$.
We can find $R_i(z)$ satisfying \eqref{cr}, whose degrees are
$(\deg R_0,\deg R_1,\deg R_{-1})=(N,N-2,N),(N,N,N-2),(N-1,N-1,N)$, etc.
In order to construct the creation/annihilation operators, however,
these minimal choices (the number of coefficients of $R_i(z)$'s is
$3N+1$) are not appropriate, because the relations
\eqref{R0cX=}--\eqref{Rm1cX='} are not satisfied, which are the desired
properties \eqref{alpmEdual}.
For this purpose, we take $(\deg R_0,\deg R_1,\deg R_{-1})=(N,N,N)$
(the number of coefficients of $R_i(z)$'s is $3N+3$).

The method of closure relation \cite{os7,os12} is the following:
(\romannumeral1) Find $R_i(z)$ satisfying \eqref{cr},
(\romannumeral2) Calculate $\alpha_{\pm}(z)$ from $R_i(z)$,
(\romannumeral3) Heisenberg solution $\underline{\mathcal{E}}_{\text{H}}(t)$
is obtained,
(\romannumeral4) Creation/ annihilation operators $a^{(\pm)}$ are obtained.
Here we change a part of the logic, namely mix (\romannumeral1)
and (\romannumeral2) by using some consequence of (\romannumeral4).
We define functions $\alpha_{\pm}(z)$ and polynomials $R_i(z)$ by
guess work, which is expected from some consequence of (\romannumeral4).
Then we check the closure relation \eqref{cr} for these $R_i(z)$.

Let us define $R_0(z)$, $R_1(z)$, $R_{-1}(z)$ and $\alpha_{\pm}(z)$ by
\begin{align}
  &R_0(z)\eqdef\sum_{j=0}^Nr_0^{(j)}z^j,\quad
  R_1(z)\eqdef\sum_{j=0}^Nr_1^{(j)}z^j,\quad
  R_{-1}(z)\eqdef\sum_{j=0}^Nr_{-1}^{(j)}z^j,
  \label{Ri}\\
  &\alpha_{\pm}(z)\eqdef
  \tfrac12\bigl(R_1(z)\pm\sqrt{R_1(z)^2+4R_0(z)}\,\bigr)
  \ \Leftrightarrow\ \left\{\!
  \begin{array}{l}
  R_0(z)=-\alpha_+(z)\alpha_-(z)\\
  R_1(z)=\alpha_+(z)+\alpha_-(z)
  \end{array}\right.,
\end{align}
where coefficients $r_0^{(j)}$ and $r_1^{(j)}$ are determined by the
condition
\begin{equation}
  \alpha_{\pm}\bigl(\check{X}(j)\bigr)
  =\check{X}(j\pm 1)-\check{X}(j)
  \ \ (j=0,1,\ldots,N),
  \label{alpmcX=}
\end{equation}
and coefficients $r_{-1}^{(j)}$ are determined by the condition
\begin{equation}
  R_{-1}\bigl(\check{X}(j)\bigr)=
  -\bigl(\check{X}(j+1)-\check{X}(j)\bigr)
  \bigl(\check{X}(j)-\check{X}(j-1)\bigr)
  B^{\text{dual}}_{\mathcal{D},j}
  \ \ (j=0,1,\ldots,N).
  \label{Rm1cX=}
\end{equation}
The condition \eqref{alpmcX=} is rewritten as
\begin{align}
  R_0\bigl(\check{X}(j)\bigr)&=
  \bigl(\check{X}(j+1)-\check{X}(j)\bigr)
  \bigl(\check{X}(j)-\check{X}(j-1)\bigr)
  \ \ (j=0,1,\ldots,N),
  \label{R0cX=}\\
  R_1\bigl(\check{X}(j)\bigr)&=
  \check{X}(j+1)-2\check{X}(j)+\check{X}(j-1)
  \ \ (j=0,1,\ldots,N),
  \label{R1cX=}
\end{align}
and the condition \eqref{Rm1cX=} is rewritten as
\begin{equation}
  R_{-1}\bigl(\check{X}(j)\bigr)=
  -R_0\bigl(\check{X}(j)\bigr)B^{\text{dual}}_{\mathcal{D},j}
  \ \ (j=0,1,\ldots,N).
  \label{Rm1cX='}
\end{equation}
Note that
\begin{equation}
  R_1\bigl(\check{X}(j)\bigr)^2+4R_0\bigl(\check{X}(j)\bigr)
  =\bigl(\check{X}(j+1)-\check{X}(j-1)\bigr)^2
  \ \ (j=0,1,\ldots,N).
  \label{R1^2+4R0}
\end{equation}
These systems of linear equations \eqref{R0cX=}, \eqref{R1cX=} and
\eqref{Rm1cX=} are solved by using Cramer's rule.
By using $\check{X}(0)=0$, we obtain
\begin{align}
  R_i(z)&=\prod_{j=1}^N\check{X}(j)^{-1}\cdot
  \prod_{1\leq k<j\leq N}\bigl(\check{X}(j)-\check{X}(k)\bigr)^{-1}\n
  &\quad\times
  \begin{vmatrix}
  \check{X}(1)&\check{X}(1)^2&\cdots&\check{X}(1)^N
  &\beta_i^{(0)}-\beta_i^{(1)}\\
  \check{X}(2)&\check{X}(2)^2&\cdots&\check{X}(2)^N
  &\beta_i^{(0)}-\beta_i^{(2)}\\
  \vdots&\vdots&\cdots&\vdots&\vdots\\
  \check{X}(N)&\check{X}(N)^2&\cdots&\check{X}(N)^N
  &\beta_i^{(0)}-\beta_i^{(N)}\\
  z&z^2&\cdots&z^N&\beta_i^{(0)}
  \end{vmatrix},
  \label{Ri2}
\end{align}
where $\beta_i^{(j)}$ ($i=0,1,-1;j=0,1,\ldots,N$) are
\begin{equation}
  \left\{\begin{array}{l}
  \beta_0^{(j)}=\bigl(\check{X}(j+1)-\check{X}(j)\bigr)
  \bigl(\check{X}(j)-\check{X}(j-1)\bigr)\\[3pt]
  \beta_1^{(j)}=\check{X}(j+1)-2\check{X}(j)+\check{X}(j-1)\\[3pt]
  \beta_{-1}^{(j)}=-\bigl(\check{X}(j+1)-\check{X}(j)\bigr)
  \bigl(\check{X}(j)-\check{X}(j-1)\bigr)
  B^{\text{dual}}_{\mathcal{D},j}
  \end{array}\right..
\end{equation}
We remark that conditions \eqref{R0cX=}, \eqref{R1cX=} and \eqref{Rm1cX=}
are $N+1$ equations for $N+1$ unknown coefficients $r_i^{(j)}$ and those
relations do not hold for $j\neq 0,1,\ldots,N$.
This contrasts with the original ($q$-)Racah systems,
in which the degrees of $R_i(z)$ are 2 or 1 and similar relations hold for
any $j$ \cite{os12}.

Then we have the following conjecture.
\begin{conj}
The closure relation \eqref{cr} holds for $R_i(z)$ \eqref{Ri}, \eqref{Ri2}.
\label{conj:cr} 
\end{conj}

{}From the closure relation \eqref{cr}, the exact Heisenberg operator
solution of the sinusoidal coordinate $\underline{\mathcal{E}}$ can be
obtained \cite{os7,os12},
\begin{align}
  \underline{\mathcal{E}}_{\text{H}}(t)
  &\eqdef e^{i\mathcal{H}^{X\,\text{dual}}_{\mathcal{D}}t}
  \underline{\mathcal{E}}\,
  e^{-i\mathcal{H}^{X\,\text{dual}}_{\mathcal{D}}t}\n
  &=a^{(+)}e^{i\alpha_+(\mathcal{H}^{X\,\text{dual}}_{\mathcal{D}})t}
  +a^{(-)}e^{i\alpha_-(\mathcal{H}^{X\,\text{dual}}_{\mathcal{D}})t}
  -R_{-1}(\mathcal{H}^{X\,\text{dual}}_{\mathcal{D}})
  R_0(\mathcal{H}^{X\,\text{dual}}_{\mathcal{D}})^{-1},
  \label{E(t)}
\end{align}
where $a^{(\pm)}=a^{(\pm)}(\mathcal{H}^{X\,\text{dual}}_{\mathcal{D}},
\underline{\mathcal{E}}\,)$ are
\begin{align}
  a^{(\pm)}&\eqdef
  \pm\Bigl([\mathcal{H}^{X\,\text{dual}}_{\mathcal{D}},
  \underline{\mathcal{E}}\,]
  -\bigl(\underline{\mathcal{E}}
  +R_{-1}(\mathcal{H}^{X\,\text{dual}}_{\mathcal{D}})
  R_0(\mathcal{H}^{X\,\text{dual}}_{\mathcal{D}})^{-1}\bigr)
  \alpha_{\mp}(\mathcal{H}^{X\,\text{dual}}_{\mathcal{D}})\Bigr)\n
  &\quad\times
  \bigl(\alpha_+(\mathcal{H}^{X\,\text{dual}}_{\mathcal{D}})
  -\alpha_-(\mathcal{H}^{X\,\text{dual}}_{\mathcal{D}})\bigr)^{-1}.
  \label{a^{(pm)}}
\end{align}
Note that square roots in
$\alpha_{\pm}(\mathcal{H}^{X\,\text{dual}}_{\mathcal{D}})$
are well-defined, because the matrix
$R_1(\mathcal{H}^{X\,\text{dual}}_{\mathcal{D}})^2
+4R_0(\mathcal{H}^{X\,\text{dual}}_{\mathcal{D}})$
is positive semi-definite, see \eqref{R1^2+4R0}.
Action of \eqref{E(t)} on $\phi^{\text{dual}}_{\mathcal{D}\,n}(x)$ is
\begin{align*}
  \underline{\mathcal{E}}_{\text{H}}(t)\phi^{\text{dual}}_{\mathcal{D}\,n}(x)
  &=e^{i\alpha_+(\mathcal{E}^{X\,\text{dual}}_{\mathcal{D},n})t}
  a^{(+)}\phi^{\text{dual}}_{\mathcal{D}\,n}(x)
  +e^{i\alpha_-(\mathcal{E}^{X\,\text{dual}}_{\mathcal{D},n})t}
  a^{(-)}\phi^{\text{dual}}_{\mathcal{D}\,n}(x)\\
  &\quad-R_{-1}(\mathcal{E}^{X\,\text{dual}}_{\mathcal{D},n})
  R_0(\mathcal{E}^{X\,\text{dual}}_{\mathcal{D},n})^{-1}
  \phi^{\text{dual}}_{\mathcal{D}\,n}(x).
\end{align*}
On the other hand it turns out to be
\begin{align*}
  &\quad\underline{\mathcal{E}}_{\text{H}}(t)
  \phi^{\text{dual}}_{\mathcal{D}\,n}(x)
  =e^{i\mathcal{H}^{X\,\text{dual}}_{\mathcal{D}}t}\underline{\mathcal{E}}\,
  e^{-i\mathcal{H}^{X\,\text{dual}}_{\mathcal{D}}t}
  \phi^{\text{dual}}_{\mathcal{D}\,n}(x)
  =e^{i\mathcal{H}^{X\,\text{dual}}_{\mathcal{D}}t}\underline{\mathcal{E}}\,
  e^{-i\mathcal{E}^{X\,\text{dual}}_{\mathcal{D},n}t}
  \phi^{\text{dual}}_{\mathcal{D}\,n}(x)\n
  &=e^{-i\mathcal{E}^{X\,\text{dual}}_{\mathcal{D},n}t}
  e^{i\mathcal{H}^{X\,\text{dual}}_{\mathcal{D}}t}\bigl(
  A^{\text{dual}}_{\mathcal{D},n}\phi^{\text{dual}}_{\mathcal{D}\,n+1}(x)
  +B^{\text{dual}}_{\mathcal{D},n}\phi^{\text{dual}}_{\mathcal{D}\,n}(x)
  +C^{\text{dual}}_{\mathcal{D},n}
  \phi^{\text{dual}}_{\mathcal{D}\,n-1}(x)\bigr)\n
  &=e^{i(\mathcal{E}^{X\,\text{dual}}_{\mathcal{D},n+1}
  -\mathcal{E}^{X\,\text{dual}}_{\mathcal{D},n})t}
  A^{\text{dual}}_{\mathcal{D},n}\phi^{\text{dual}}_{\mathcal{D}\,n+1}(x)
  +B^{\text{dual}}_{\mathcal{D},n}\phi^{\text{dual}}_{\mathcal{D}\,n}(x)
  +e^{i(\mathcal{E}^{X\,\text{dual}}_{\mathcal{D},n-1}
  -\mathcal{E}^{X\,\text{dual}}_{\mathcal{D},n})t}
  C^{\text{dual}}_{\mathcal{D},n}\phi^{\text{dual}}_{\mathcal{D}\,n-1}(x),
\end{align*}
where we have used \eqref{dualphiDn3term}.
Comparing these $t$-dependence, we obtain
\begin{align}
  &\alpha_{\pm}(\mathcal{E}^{X\,\text{dual}}_{\mathcal{D},n})
  =\mathcal{E}^{X\,\text{dual}}_{\mathcal{D},n\pm 1}
  -\mathcal{E}^{X\,\text{dual}}_{\mathcal{D},n},\quad
  -R_{-1}(\mathcal{E}^{X\,\text{dual}}_{\mathcal{D},n})
  R_0(\mathcal{E}^{X\,\text{dual}}_{\mathcal{D},n})^{-1}
  =B^{\text{dual}}_{\mathcal{D},n},
  \label{alpmEdual}\\
  &a^{(+)}\phi^{\text{dual}}_{\mathcal{D}\,n}(x)
  =A^{\text{dual}}_{\mathcal{D},n}
  \phi^{\text{dual}}_{\mathcal{D}\,n+1}(x),\quad
  a^{(-)}\phi^{\text{dual}}_{\mathcal{D}\,n}(x)
  =C^{\text{dual}}_{\mathcal{D},n}\phi^{\text{dual}}_{\mathcal{D}\,n-1}(x).
  \label{act_a+a-}
\end{align}
Therefore $a^{(+)}$ and $a^{(-)}$ are creation and annihilation operators,
respectively.
The relations \eqref{alpmEdual} correspond to \eqref{R0cX=}--\eqref{Rm1cX='}.

\noindent
{\bf Remark:} The value of function $\check{X}(x)$ \eqref{cX}
at $x=-1$ is
\begin{equation}
  \check{X}(-1)=\left\{
  \begin{array}{ll}
  -(d+M-1)Y(0)&:\text{R}\\[2pt]
  -(1-q)(1-dq^{M-1})Y(0)&:\text{$q$R}
  \end{array}\right..
\end{equation}
For $Y(0)=0$, we have $\check{X}(-1)=0$. By \eqref{R0cX=}, this and
$\check{X}(0)=0$ give $r_0^{(0)}=0$, namely $R_0(0)=0$.
Therefore action of $a^{(\pm)}$ on $\phi^{\text{dual}}_{\mathcal{D}\,0}(x)$
is not well-defined for this case.
Although the coefficient $r_{-1}^{(0)}$ also vanishes by \eqref{Rm1cX=},
namely $R_{-1}(0)=0$, and the limit $\lim\limits_{z\to 0}R_{-1}(z)R_0(z)^{-1}$
exists, it does not coincide with $-B^{\text{dual}}_{\mathcal{D},0}$.

By the similarity transformation, the closure relation \eqref{cr} is
rewritten for the similarity transformed Hamiltonian 
$\widetilde{\mathcal{H}}^{X\,\text{dual}}_{\mathcal{D}}(\bm{\lambda})$
\eqref{dualtH},
\begin{equation}
  \bigl[\widetilde{\mathcal{H}}^{X\,\text{dual}}_{\mathcal{D}},
  [\widetilde{\mathcal{H}}^{X\,\text{dual}}_{\mathcal{D}},
  \underline{\mathcal{E}}\,]\bigr]
  =\underline{\mathcal{E}}\,
  R_0(\widetilde{\mathcal{H}}^{X\,\text{dual}}_{\mathcal{D}})
  +[\widetilde{\mathcal{H}}^{X\,\text{dual}}_{\mathcal{D}},
  \underline{\mathcal{E}}\,]
  R_1(\widetilde{\mathcal{H}}^{X\,\text{dual}}_{\mathcal{D}})
  +R_{-1}(\widetilde{\mathcal{H}}^{X\,\text{dual}}_{\mathcal{D}}).
  \label{tcr}
\end{equation}
{}From the creation and annihilation operators
$a^{(\pm)}=a^{(\pm)}(\mathcal{H}^{X\,\text{dual}}_{\mathcal{D}},
\underline{\mathcal{E}}\,)$ \eqref{a^{(pm)}},
we obtain the creation and annihilation operators for polynomial eigenvectors,
\begin{align}
  &\tilde{a}^{(\pm)}\eqdef
  \phi^{\text{dual}}_{\mathcal{D}\,0}(x)^{-1}\circ
  a^{(\pm)}(\mathcal{H}^{X\,\text{dual}}_{\mathcal{D}},
  \underline{\mathcal{E}}\,)\circ
  \phi^{\text{dual}}_{\mathcal{D}\,0}(x)
  =a^{(\pm)}(\widetilde{\mathcal{H}}^{X\,\text{dual}}_{\mathcal{D}},
  \underline{\mathcal{E}}\,),\\
  &\tilde{a}^{(+)}\check{Q}_{\mathcal{D},n}(x)
  =A^{\text{dual}}_{\mathcal{D},n}\check{Q}_{\mathcal{D},n+1}(x),\quad
  \tilde{a}^{(-)}\check{Q}_{\mathcal{D},n}(x)
  =C^{\text{dual}}_{\mathcal{D},n}\check{Q}_{\mathcal{D},n-1}(x).
  \label{act_ta+ta-}
\end{align}

\subsection{No shape invariance}
\label{sec:si}

We will show that the rdQM system described by
$\mathcal{H}^{X\,\text{dual}}_{\mathcal{D}}(\bm{\lambda})$ is not shape
invariant.

First let us factorize the Hamiltonian
$\mathcal{H}^{X\,\text{dual}}_{\mathcal{D}}(\bm{\lambda})$ \eqref{HDXdual}.
Since it is positive semi-definite, it can be factorized as
\begin{equation}
  \mathcal{H}^{X\,\text{dual}}_{\mathcal{D}}(\bm{\lambda})
  =\mathcal{A}^{X\,\text{dual}}_{\mathcal{D}}(\bm{\lambda})^{\dagger}
  \mathcal{A}^{X\,\text{dual}}_{\mathcal{D}}(\bm{\lambda}),
\end{equation}
where $\mathcal{A}^{X\,\text{dual}}_{\mathcal{D}}(\bm{\lambda})$ is an
upper triangular matrix (with upper bandwidth $L$ for $L\leq\frac12N$).
By imposing the condition
$\mathcal{A}^{X\,\text{dual}}_{\mathcal{D}}(\bm{\lambda})_{x,x}\geq 0$
$(x=0,1,\ldots,x_{\text{max}})$, this upper triangular matrix
$\mathcal{A}^{X\,\text{dual}}_{\mathcal{D}}(\bm{\lambda})
=(a_{x,y})_{0\leq x,y\leq x_{\text{max}}}$ is given by
\begin{align}
  a_{x,y}=\left\{
  \begin{array}{ll}
  0&(0\leq y\leq x-1)\\[2pt]
  \sqrt{h'_{x,x}}&(y=x)\\[2pt]
  {\displaystyle\frac{h'_{x,y}}{\sqrt{h'_{x,x}}}}
  &(x+1\leq y\leq x_{\text{max}})
  \end{array}\right.\quad
  (0\leq x\leq x_{\text{max}}).
\end{align}
Here $h'_{x,y}$ are defined by
\begin{equation}
  h'_{x,y}=h_{x,y}-\sum_{z=0}^{x-1}\frac{h'_{z,x}h'_{z,y}}{\sqrt{h'_{z,z}}}
  \ \ (0\leq x\leq y\leq x_{\text{max}}),
\end{equation}
where $h_{x,y}=\mathcal{H}^{X\,\text{dual}}_{\mathcal{D}}(\bm{\lambda})_{x,y}$
and the convention $\sum_{i=n}^{n-1}*=0$ is assumed.
Note that the zero eigenvalue of
$\mathcal{H}^{X\,\text{dual}}_{\mathcal{D}}(\bm{\lambda})$ implies
$\mathcal{A}^{X\,\text{dual}}_{\mathcal{D}}
(\bm{\lambda})_{x_{\text{max}},x_{\text{max}}}=0$.
So, the last row of $\mathcal{A}^{X\,\text{dual}}_{\mathcal{D}}(\bm{\lambda})$
is zero.

Next we recall the general theory of the shape invariance for finite rdQM
systems ($x_{\text{max}}=n_{\text{max}}=N$) \cite{os12}.
The Hamiltonian $\mathcal{H}(\bm{\lambda})=
(\mathcal{H}(\bm{\lambda})_{x,y})_{0\leq x,y\leq x_{\text{max}}}$ is positive
semi-definite, whose eigenvalues are
$0=\mathcal{E}_0(\bm{\lambda})<\mathcal{E}_1(\bm{\lambda})<\cdots<
\mathcal{E}_{n_{\text{max}}}(\bm{\lambda})$ and corresponding eigenvectors
are $\phi_n(x;\bm{\lambda})$, and factorized as
$\mathcal{H}(\bm{\lambda})=\mathcal{A}(\bm{\lambda})^{\dagger}
\mathcal{A}(\bm{\lambda})$, where $\mathcal{A}(\bm{\lambda})$ is upper
triangular and $\mathcal{A}(\bm{\lambda})_{x_{\text{max}},x_{\text{max}}}=0$.
Since the last row of $\mathcal{A}(\bm{\lambda})$ is zero, we have
\begin{equation*}
  \mathcal{A}(\bm{\lambda})\mathcal{A}(\bm{\lambda})^{\dagger}=
  \begin{pmatrix}
  B&\vec{0}\,\\
  {}^t\vec{0}&0
  \end{pmatrix},\quad
  \mathcal{A}(\bm{\lambda})\phi_n(x;\bm{\lambda})
  =\begin{pmatrix}
  \vec{b}\,\\0
  \end{pmatrix}.
\end{equation*}
Let us write these $N\times N$ matrix $B$ and $N$ component vector $\vec{b}$
as
\begin{equation}
  B=\bigl(\mathcal{A}(\bm{\lambda})\mathcal{A}(\bm{\lambda})^{\dagger}
  \bigr)^{[N\times N]},\quad
  \vec{b}=\bigl(\mathcal{A}(\bm{\lambda})\phi_n(x;\bm{\lambda})\bigr)^{[N]}.
\end{equation}
(In our previous studies \cite{os12}--\cite{os34}, we treat tridiagonal
Hamiltonians and $\mathcal{A}$ is an upper triangular matrix with upper
bandwidth 1. Here this is not assumed.)
Shape invariance is a relation between the system with parameters
$\bm{\lambda}$ and that with $\bm{\lambda'}$. Usually, appropriate choice of
parameters allow us to express $\bm{\lambda'}$ as shifts of parameters
$\bm{\lambda'}=\bm{\lambda}+\bm{\delta}$, but here we do not assume this.
The number $N$, which corresponds to the size of the Hamiltonian, is one
element of $\bm{\lambda}$ and it changes to $N-1$ in $\bm{\lambda'}$.
Then the shape invariant condition is
\begin{equation}
  \bigl(\mathcal{A}(\bm{\lambda})\mathcal{A}(\bm{\lambda})^{\dagger}
  \bigr)^{[N\times N]}
  =\kappa\mathcal{A}(\bm{\lambda'})^{\dagger}\mathcal{A}(\bm{\lambda'})
  +\mathcal{E}_1(\bm{\lambda}),
  \label{si}
\end{equation}
where $\kappa$ is a positive constant and $\mathcal{E}_1(\bm{\lambda})$
is the abbreviation for $\mathcal{E}_1(\bm{\lambda})\bm{1}_N$.
The Darboux transformation is defined by
\begin{align}
  \mathcal{H}^{\text{new}}(\bm{\lambda})&\eqdef
  \mathcal{A}(\bm{\lambda'})^{\dagger}\mathcal{A}(\bm{\lambda'})
  =\mathcal{H}(\bm{\lambda'}),\\
  \phi^{\text{new}}_n(x;\bm{\lambda})&\eqdef
  \bigl(\mathcal{A}(\bm{\lambda})\phi_{n+1}(x;\bm{\lambda})\bigr)^{[N]}
  \ \ (n=0,1,\ldots,N-1).
\end{align}
The shape invariant condition \eqref{si} gives
\begin{align*}
  &\quad
  \bigl(\mathcal{A}(\bm{\lambda})\mathcal{A}(\bm{\lambda})^{\dagger}
  \bigr)^{[N\times N]}\phi^{\text{new}}_n(x;\bm{\lambda})
  =\bigl(\kappa\mathcal{A}(\bm{\lambda'})^{\dagger}\mathcal{A}(\bm{\lambda'})
  +\mathcal{E}_1(\bm{\lambda})\bigr)\phi^{\text{new}}_n(x;\bm{\lambda})\\
  &=\left(
  \begin{pmatrix}
  \bigl(\mathcal{A}(\bm{\lambda})\mathcal{A}(\bm{\lambda})^{\dagger}
  \bigr)^{[N\times N]}&\vec{0}\\
  {}^t\vec{0}&0
  \end{pmatrix}
  \begin{pmatrix}
  \bigl(\mathcal{A}(\bm{\lambda})\phi_{n+1}(x;\bm{\lambda})\bigr)^{[N]}\\0
  \end{pmatrix}\right)^{[N]}\\
  &=\Bigl(\mathcal{A}(\bm{\lambda})\mathcal{A}(\bm{\lambda})^{\dagger}\cdot
  \mathcal{A}(\bm{\lambda})\phi_{n+1}(x;\bm{\lambda})\Bigr)^{[N]}\\
  &=\Bigl(\mathcal{A}(\bm{\lambda})\cdot\mathcal{A}(\bm{\lambda})^{\dagger}
  \mathcal{A}(\bm{\lambda})\phi_{n+1}(x;\bm{\lambda})\Bigr)^{[N]}
  =\Bigl(\mathcal{A}(\bm{\lambda})\cdot\mathcal{H}(\bm{\lambda})
  \phi_{n+1}(x;\bm{\lambda})\Bigr)^{[N]}\\
  &=\Bigl(\mathcal{E}_{n+1}(\bm{\lambda})\mathcal{A}(\bm{\lambda})
  \phi_{n+1}(x;\bm{\lambda})\Bigr)^{[N]}
  =\mathcal{E}_{n+1}(\bm{\lambda})\phi^{\text{new}}_n(x;\bm{\lambda}),
\end{align*}
namely,
\begin{equation}
  \mathcal{H}^{\text{new}}(\bm{\lambda})\phi^{\text{new}}_n(x;\bm{\lambda})
  =\frac{1}{\kappa}\bigl(\mathcal{E}_{n+1}(\bm{\lambda})
  -\mathcal{E}_1(\bm{\lambda})\bigr)\phi^{\text{new}}_n(x;\bm{\lambda}).
\end{equation}
{}From the relation
$\mathcal{H}^{\text{new}}(\bm{\lambda})=\mathcal{H}(\bm{\lambda'})$,
we obtain
\begin{equation}
  \mathcal{E}_n(\bm{\lambda'})
  =\frac{1}{\kappa}\bigl(\mathcal{E}_{n+1}(\bm{\lambda})
  -\mathcal{E}_1(\bm{\lambda})\bigr)
  \ \ (n=0,1,\ldots,N-1).
  \label{En(la')}
\end{equation}
This relation implies that the energy eigenvalues $\mathcal{E}_n(\bm{\lambda})$
are determined by the information of the first excited state energy
$\mathcal{E}_1(\bm{\lambda})$.
For example, the energy eigenvalues $\mathcal{E}_n(\bm{\lambda})$ for the
original ($q$-)Racah systems are given by \eqref{En}, and new set of
parameters $\bm{\lambda'}$ is $\bm{\lambda}+\bm{\delta}$.
It is easy to check \eqref{En(la')} and
$\mathcal{E}_n(\bm{\lambda})=\sum\limits_{s=0}^{n-1}\kappa^s
\mathcal{E}_1(\bm{\lambda}+s\bm{\delta})$ ($n=0,1,\ldots,N$) for these cases.

Now let us consider the shape invariance for the new exactly solvable rdQM
$\mathcal{H}^{X\,\text{dual}}_{\mathcal{D}}(\bm{\lambda})$.
Assume that this system is shape invariant,
\begin{equation*}
  \bigl(\mathcal{A}^{X\,\text{dual}}_{\mathcal{D}}(\bm{\lambda})
  \mathcal{A}^{X\,\text{dual}}_{\mathcal{D}}(\bm{\lambda})^{\dagger}
  \bigr)^{[N\times N]}
  =\kappa^{\text{dual}}
  \mathcal{A}^{X\,\text{dual}}_{\mathcal{D}}(\bm{\lambda'})^{\dagger}
  \mathcal{A}^{X\,\text{dual}}_{\mathcal{D}}(\bm{\lambda'})
  +\mathcal{E}^{X\,\text{dual}}_{\mathcal{D},1}(\bm{\lambda}).
\end{equation*}
Then \eqref{En(la')} gives
\begin{equation*}
  \mathcal{E}^{X\,\text{dual}}_{\mathcal{D},n}(\bm{\lambda'})
  =\frac{1}{\kappa^{\text{dual}}}
  \bigl(\mathcal{E}^{X\,\text{dual}}_{\mathcal{D},n+1}(\bm{\lambda})
  -\mathcal{E}^{X\,\text{dual}}_{\mathcal{D},1}(\bm{\lambda})\bigr)
  \ \ (n=0,1,\ldots,N-1),
\end{equation*}
namely,
\begin{equation*}
  \check{X}(x;\bm{\lambda'})
  =\frac{1}{\kappa^{\text{dual}}}
  \bigl(\check{X}(x+1;\bm{\lambda})-\check{X}(1;\bm{\lambda})\bigr)
  \ \ (x=0,1,\ldots,N-1).
\end{equation*}
However, these relations do not hold in general, because the concrete
expression of $\mathcal{E}^{X\,\text{dual}}_{\mathcal{D},n}(\bm{\lambda})
=\check{X}(n;\bm{\lambda})$ is much more complicated than
$\mathcal{E}_n(\bm{\lambda})$ \eqref{En}.
We will convince this by calculating small $M$ examples, see \S\,\ref{sec:ex}.
Therefore the new exactly solvable rdQM
$\mathcal{H}^{X\,\text{dual}}_{\mathcal{D}}(\bm{\lambda})$ is not
shape invariant.

\subsection{Examples}
\label{sec:ex}

To write down the Hamiltonian
$\mathcal{H}^{X\,\text{dual}}_{\mathcal{D}}(\bm{\lambda})$ \eqref{HDXdual}
or similarity transformed one
$\widetilde{\mathcal{H}}^{X\,\text{dual}}_{\mathcal{D}}(\bm{\lambda})$
\eqref{dualtH}, we need explicit form of the coefficients
$r_{n,k}^{X,\mathcal{D}}(\bm{\lambda})$ \eqref{cXcP}.
We can calculate $r_{n,k}^{X,\mathcal{D}}$ explicitly for small $M$, $d_j$,
$\deg Y$ and $n$, and check various properties of the system for small $N$:
the Schr\"odinger equation \eqref{dualHphiDn=} (or \eqref{dualtHQDn=}),
commutativity \eqref{[HX1,HX2]=0}, the closure relation \eqref{cr}
(or \eqref{tcr}), action of the creation/annihilation operators
\eqref{act_a+a-} (or \eqref{act_ta+ta-}), etc.
Monotonically increasing property of the eigenvalues \eqref{Eincrease}
can be also checked for
small $N$ by assigning various numerical values to $q,b,c,d$ satisfying
\eqref{para}.
However, to find the closed expression of $r_{n,k}^{X,\mathcal{D}}$ for
general $n$ is very difficult.
We have obtained such general $n$ expression of $r_{n,k}^{X,\mathcal{D}}$ for
\begin{align*}
  \text{R}:\ \ &\mathcal{D}=\{1\},Y=1;
  \ \mathcal{D}=\{2\},Y=1;
  \ \mathcal{D}=\{1,2\},Y=1;
  \ \mathcal{D}=\{1\},Y(\eta)=\eta,\\
  \text{$q$R}:\ \ &\mathcal{D}=\{1\},Y=1;
  \ \mathcal{D}=\{2\},Y=1,
\end{align*}
but their explicit forms are somewhat lengthy.
Here we write down $r_{n,k}^{X,\mathcal{D}}$ for $\mathcal{D}=\{1\}$ and
$Y=1$ \cite{rrmiop5}.
For other cases, we present $X(\eta)$ only.
{}From $X(\eta)=X(\eta;\bm{\lambda})$ and \eqref{cX}, the energy eigenvalues 
$\mathcal{E}^{X\,\text{dual}}_{\mathcal{D},n}(\bm{\lambda})$ \eqref{EXdualDn}
are obtained.
Since the overall normalization of $X(\eta)$ is not important,
we multiply $X(\eta)$ \eqref{X=I[XiY]} by an appropriate positive factor.

\subsubsection{dual multi-indexed Racah systems}
\label{sec:ex_R}

We set $\sigma_1=a+b$, $\sigma_2=ab$, $\sigma'_1=c+d$ and $\sigma'_2=cd$.\\
\noindent
\underline{Ex.1} $\mathcal{D}=\{1\}$, $Y(\eta)=1$
($\Rightarrow$ $L=2$, $X=X_{\text{min}}$)
\begin{align}
  X(\eta)&=-2c(d-a+1)(d-b+1)
  I_{\bm{\lambda}+\bm{\delta}}[\Xi_{\mathcal{D}}](\eta)\n
  &=-\eta\bigl((2-\sigma_1+\sigma'_1)\eta
  -\sigma_1(2c+d+2\sigma'_2)+2\sigma_2c+2\sigma'_1
  +\sigma'_2(5+2d)+d^2\bigr),\\
  r_{n,2}^{X,\mathcal{D}}&=
  -\frac{(2-\sigma_1+\sigma'_1)(c+n)(c+n+3)(a+n,b+n,\tilde{d}+n)_2}
  {(\tilde{d}+2n)_4},\n
  r_{n,-2}^{X,\mathcal{D}}&=
  -\frac{(2-\sigma_1+\sigma'_1)(\tilde{d}-c+n-3)(\tilde{d}-c+n)
  (\tilde{d}-a+n-1,\tilde{d}-b+n-1,n-1)_2}
  {(\tilde{d}+2n-3)_4},\n
  r_{n,1}^{X,\mathcal{D}}&=
  -\frac{2(a+n)(b+n)(c+n)(c+n+2)(\tilde{d}-c+n)(\tilde{d}+n)}
  {(\tilde{d}+2n+3)(\tilde{d}+2n-1)_3}\n
  &\quad\times
  \Bigl(-2(2-\sigma_1+\sigma'_1)n(n+\tilde{d}+1)
  +2(1-\tilde{d})(1+c-\sigma_2)+d(1-\tilde{d}^2)\Bigr),\\
  r_{n,-1}^{X,\mathcal{D}}&=
  -\frac{2n(\tilde{d}-a+n)(\tilde{d}-b+n)(c+n)(\tilde{d}-c+n-2)(\tilde{d}-c+n)}
  {(\tilde{d}+2n-3)(\tilde{d}+2n-1)_3}\n
  &\quad\times
  \Bigl(-2(2-\sigma_1+\sigma'_1)n(n+\tilde{d}-1)
  +2(1+c-\sigma_2)+2(\sigma_2+c-\tilde{d})\tilde{d}+d(1-\tilde{d}^2)\Bigr),\n
  r_{n,0}^{X,\mathcal{D}}&=
  -\sum_{k=1}^2\bigl(r_{n,-k}^{X,\mathcal{D}}+r_{n,k}^{X,\mathcal{D}}\bigr).
  \nonumber
\end{align}
\underline{Ex.2} $\mathcal{D}=\{2\}$, $Y(\eta)=1$
($\Rightarrow$ $L=3$, $X=X_{\text{min}}$)
\begin{align}
  X(\eta)&=3c(1+c)(d-a+1)(d-a+2)(d-b+1)(d-b+2)
  I_{\bm{\lambda}+\bm{\delta}}[\Xi_{\mathcal{D}}](\eta)\n
  &=\eta\Bigl(
  (\sigma_1-\sigma'_1-4)(\sigma_1-\sigma'_1-3)\eta^2\n
  &\qquad
  -(\sigma_1-\sigma'_1-3)\bigl(3(1+c)(d-a)(d-b)+2(5+6c)d
  -2\sigma_1(2+3c)+4+10c\bigr)\eta\n
  &\qquad
  +(3c^2+6c+2)d^2(d-\sigma_1)^2+(12+40c+21c^2)d^3\\
  &\qquad
  +\bigl(22+88c+50c^2-\sigma_1(16+55c+30c^2)+\sigma_2(3+9c+6c^2)\bigr)d^2\n
  &\qquad
  +\bigl(12+70c+46c^2-\sigma_1(16+71c+45c^2)+\sigma_1^2(4+15c+9c^2)
  +3\sigma_2(3+10c+7c^2)\n
  &\qquad\quad
  -3\sigma_1\sigma_2(1+c)(1+2c)\bigr)d
  +3(a-2)(a-1)(b-2)(b-1)c(c+1)\Bigr).\nonumber
\end{align}
\underline{Ex.3} $\mathcal{D}=\{1,2\}$, $Y(\eta)=1$
($\Rightarrow$ $L=3$, $X=X_{\text{min}}$)
\begin{align}
  X(\eta)&=3c(c+1)(d-a+1)(d-a+2)(d-b+1)(d-b+2)
  I_{\bm{\lambda}+2\bm{\delta}}[\Xi_{\mathcal{D}}](\eta)\n
  &=\eta\Bigl(
  (\sigma_1-\sigma'_1-3)(\sigma_1-\sigma'_1-2)\eta^2\n
  &\qquad
  -(\sigma_1-\sigma'_1-3)\bigl(3(1+c)d(d-\sigma_1)+(7+9c)d+2+4c-\sigma_1
  -3c(\sigma_1-\sigma_2)\bigr)\eta\n
  &\qquad
  +(2+6c+3c^2)d^2(d-\sigma_1)^2+(12+40c+21c^2)d^3\\
  &\qquad
  +\bigl(22+89c+50c^2-\sigma_1(14+55c+30c^2)+3\sigma_2c(3+2c)\bigr)d^2\n
  &\qquad
  +\bigl(12+76c+47c^2-\sigma_1(10+73c+45c^2)+\sigma_1^2(2+15c+9c^2)
  -3\sigma_1\sigma_2c(3+2c)\n
  &\qquad\quad
  +3\sigma_2c(10+7c)\bigr)d
  -3(\sigma_1-\sigma_2-1)c\bigl(7+5c-\sigma_1(3+2c)+\sigma_2(1+c)\bigr)
  \Bigr).\nonumber
\end{align}
\underline{Ex.4} $\mathcal{D}=\{1\}$, $Y(\eta)=\eta$ ($\Rightarrow$ $L=3$)
\begin{align}
  X(\eta)&=-6c(d-a+1)(d-b+1)
  I_{\bm{\lambda}+\bm{\delta}}[\Xi_{\mathcal{D}}Y](\eta)\n
  &=-\eta\Bigl(
  2(\sigma'_1-\sigma_1+2)\eta^2\n
  &\qquad\quad+\bigl(3d(1+c)(d-\sigma_1)+(5+9c)d-2+2c+\sigma_1
  +3c(\sigma_2-\sigma_1)\bigr)\eta\\
  &\qquad\quad
  +d\bigl(d(1+3c)(d-\sigma_1)+(1+7c)d-2+2c+\sigma_1
  +3c(\sigma_2-\sigma_1)\bigr)\Bigr).\nonumber
\end{align}

\subsubsection{dual multi-indexed $q$-Racah systems}
\label{sec:ex_qR}

We set $\sigma_1=a+b$, $\sigma_2=ab$, $\sigma'_1=c+d$ and $\sigma'_2=cd$.\\
\noindent
\underline{Ex.1} $\mathcal{D}=\{1\}$, $Y(\eta)=1$
($\Rightarrow$ $L=2$, $X=X_{\text{min}}$)
\begin{align}
  X(\eta)&=-(1+q)(1-c)(1-a^{-1}dq)(1-b^{-1}dq)
  I_{\bm{\lambda}+\bm{\delta}}[\Xi_{\mathcal{D}}](\eta)\n
  &=-\eta\Bigl((1-\sigma_2^{-1}\sigma'_2q^2)\eta
  +\sigma_2^{-1}q^2(1+q-2cq)d^2\n
  &\phantom{\quad-\eta\Bigl(}
  -\sigma_2^{-1}\bigl(\sigma_1q(1+q)(1-c)+(1-q)(\sigma_2+cq^2)\bigr)d
  +2-c(1+q)\Bigr),\\
  r_{n,2}^{X,\mathcal{D}}&=
  -\frac{(1-\sigma_2^{-1}\sigma'_2q^2)(1-cq^n)(1-cq^{n+3})
  (aq^n,bq^n,\tilde{d}q^n;q)_2}
  {(\tilde{d}q^{2n};q)_4},\n
  r_{n,-2}^{X,\mathcal{D}}&=
  -\frac{d^2q^2(1-\sigma_2^{-1}\sigma'_2q^2)
  (1-c^{-1}\tilde{d}q^{n-3})(1-c^{-1}\tilde{d}q^n)
  (a^{-1}\tilde{d}q^{n-1},b^{-1}\tilde{d}q^{n-1},q^{n-1};q)_2}
  {(\tilde{d}q^{2n-3};q)_4},\n
  r_{n,1}^{X,\mathcal{D}}&=
  -\frac{(1+q)(1-aq^n)(1-bq^n)(1-cq^n)(1-cq^{n+2})(1-c^{-1}\tilde{d}q^n)
  (1-\tilde{d}q^n)}
  {\sigma_2d
  (1-\tilde{d}q^{2n+3})(\tilde{d}q^{2n-1};q)_3}\n
  &\quad\times
  \Bigl(-\bigl(\sigma_2\sigma'_1+\sigma_1(1-c)dq
  -\sigma'_1dq^2\bigr)(\sigma_2cq^{2n}+d)\n
  &\phantom{\quad\times\Bigl(}
  +(q+q^{-1})d\bigl(\sigma_1\sigma_2c+\sigma_2(1-c)\sigma'_1q
  -\sigma_1\sigma'_2q^2\bigr)q^n\Bigr),\\
  r_{n,-1}^{X,\mathcal{D}}&=
  -\frac{(1+q)(1-q^n)(1-a^{-1}\tilde{d}q^n)(1-b^{-1}\tilde{d}q^n)
  (1-cq^n)(1-c^{-1}\tilde{d}q^{n-2})(1-c^{-1}\tilde{d}q^n)}
  {\sigma_2
  (1-\tilde{d}q^{2n-3})(\tilde{d}q^{2n-1};q)_3}\n
  &\quad\times
  \Bigl(-\bigl(\sigma_2\sigma'_1
  +\sigma_1(1-c)dq-\sigma'_1dq^2\bigr)(\sigma_2cq^{2n-1}+dq)\n
  &\phantom{\quad\times\Bigl(}
  +(q+q^{-1})d\bigl(\sigma_1\sigma_2c+\sigma_2(1-c)\sigma'_1q
  -\sigma_1\sigma'_2q^2\bigr)q^n
  \Bigr),\n
  r_{n,0}^{X,\mathcal{D}}&=
  -\sum_{k=1}^2\bigl(r_{n,-k}^{X,\mathcal{D}}+r_{n,k}^{X,\mathcal{D}}\bigr).
  \nonumber
\end{align}
\underline{Ex.2} $\mathcal{D}=\{2\}$, $Y(\eta)=1$
($\Rightarrow$ $L=3$, $X=X_{\text{min}}$)
\begin{align}
  X(\eta)&
  =(1+q+q^2)(1-c)(1-c q)(a-d q)(a-d q^2)(b-d q)(b-d q^2)
  I_{\bm{\lambda}+\bm{\delta}}[\Xi_{\mathcal{D}}](\eta)\n
  &=\eta\Bigl(
  (\sigma_2-cdq^3)(\sigma_2-cdq^4)\eta^2
  +(\sigma_2-cdq^3)\bigl((1+q+q^2)(q^3d^2-q(1-cq)\sigma_1d-c\sigma_2)\n
  &\qquad\quad
  -3cq^5d^2-(1-q)^2(\sigma_2-cq^3)d+3\sigma_2\bigr)\eta\n
  &\qquad
  +(1+q+q^2)\Bigl(q^6(1-2cq)d^4
  -q^3\bigl(q(1-cq)(1+q-2cq)\sigma_1+(1-q)(\sigma_2+c q^3)\bigr)d^3\n
  &\qquad\quad
  +q\bigl(q^2(1-c)(1-cq)\sigma_1^2+(1-q)(1-cq)(\sigma_2+cq^3)\sigma_1
  +q(1+q)(1+c^2 q^2)\sigma_2\bigr)d^2\n
  &\qquad\quad
  -\bigl(q(1-cq)(2-(1+q)c)\sigma_1-(1-q)(\sigma_2+q^3 c)c\bigr)\sigma_2d
  -(2-cq)c\sigma_2^2\Bigr)\\
  &\qquad
   +3q^9 c^2 d^4
   +q^4(1-q)\bigl((2+q)\sigma_2+q^2(1+2q^2)c\bigr)c d^3\n
  &\qquad
  -q\bigl((1-q)^2(\sigma_2^2+q^5c^2)+q(1+q)(1+4q^2+q^4)c\sigma_2\bigr)d^2\n
  &\qquad
  -\sigma_2(1-q)\bigl((2+q^2)\sigma_2+q^3(1+2q)c\bigr)d
   +3\sigma_2^2\Bigr).\nonumber
\end{align}

\section{Summary and Comments}
\label{sec:summary}

The case-(1) multi-indexed ($M$-indexed) ($q$-)Racah orthogonal polynomials
$\check{P}_{\mathcal{D},n}(x)$ satisfy the second order difference equations
\eqref{tHDcPDn=} \cite{os26} and various
$1+2L$ ($L\geq M+1$) term recurrence relations with constant coefficients
\eqref{cXcP} \cite{rrmiop5}.
Corresponding to these properties, their dual polynomials
$\check{Q}_{\mathcal{D},x}(n)$ \eqref{QDxn} satisfy the three term recurrence
relations \eqref{3termcQ} and various $2L$-th order difference equations
\eqref{XQ}.
That is, the dual multi-indexed ($q$-)Racah polynomials are ordinary
orthogonal polynomials and the Krall-type.
Their weight functions do not contain delta functions (Kronecker deltas).

We construct new exactly solvable rdQM systems, whose eigenvectors are
described by the dual multi-indexed ($q$-)Racah polynomials. Their
Hamiltonians \eqref{HDXdual} are not tridiagonal but ``$(1+2L)$-diagonal''.
These quantum systems satisfy the closure relations \eqref{cr}, from which
the creation/annihilation operators \eqref{act_a+a-} are obtained, but they
are not shape invariant.
As a sufficient condition for exact solvability, we know two conditions:
the closure relation and the shape invariance.
Concerning the exactly solvable models we have studied,
we observe that when they satisfy the (generalized) closure relation,
they are also shape invariant.
The new exactly solvable rdQM systems \eqref{HDXdual} give counterexamples
to this observation.

Finally we list some problems related to the dual multi-indexed ($q$-)Racah
polynomials.
\begin{enumerate}
\item
The commutativity \eqref{[HX1,HX2]=0} originates from the non-uniqueness
of $X$ giving the recurrence relations with constant coefficients \eqref{cXcP}.
The relations among the recurrence relations for various $X$ ($Y$) are
unclear. It is an important problem to clarify them.
\item
Orthogonal polynomials of a discrete variable in the Askey-scheme can
be obtained as certain limits of the ($q$-)Racah polynomials \cite{kls}.
It is an interesting problem to study various limits of the (dual)
multi-indexed ($q$-)Racah polynomials.
We remark that the (dual) multi-indexed ($q$-)Racah polynomials may not
reduce to good polynomials in the same limits used for the ($q$-)Racah
polynomials. For example, the case-(1) multi-indexed polynomials are not
allowed for some reduced polynomials.
See \cite{d16} for similar situation in the Krall-type case.
\item
For each (exactly solvable) rdQM system, we can construct the (exactly
solvable) birth and death process \cite{s09}, which is a stationary Markov
chain. The new exactly solvable systems described by the dual multi-indexed
($q$-)Racah polynomials provide new exactly solvable birth and death processes.
\item
We may be able to deform the new exactly solvable systems \eqref{HDXdual}
by multi-step Darboux transformations with appropriate seed solutions.
At least it is possible to take eigenvectors as seed solutions.
However, some formulas in \cite{os22} may be modified, because the
Hamiltonians \eqref{HDXdual} are not tridiagonal.
It is an interesting problem to study such deformation and to clarify whether
virtual state vectors exist or not.
This will give examples of the combined three directions
(\romannumeral1)--(\romannumeral3) in \S\,\ref{sec:intro}.
\item
The case-(1) multi-indexed polynomials of the Laguerre, Jacobi, Wilson and
Askey-Wilson types satisfy the second order difference equations
\cite{os25,os27} and various $1+2L$ term recurrence relations with constant
coefficients \cite{rrmiop2,rrmiop3}. But their variable $x$ is continuous
and dual polynomials are not defined naturally.
It is a challenging problem to construct the Krall-type polynomials related
to these multi-indexed polynomials.
\end{enumerate}

\section*{Acknowledgments}

I thank R.\,Sasaki for discussion and useful comments on the manuscript.

\bigskip
\appendix
\section{Data for Multi-indexed ($q$-)Racah Polynomials}
\label{app:data}

For readers' convenience, we present some data for the multi-indexed
($q$-)Racah polynomials \cite{os12,os26,rrmiop5},
which are not presented in the main text.

\noindent
$\bullet$ ($q$-)Racah polynomials $P_n(\eta;\bm{\lambda})$:
\begin{align}
  &\check{P}_n(x;\bm{\lambda})
  \eqdef P_n\bigl(\eta(x;\bm{\lambda});\bm{\lambda}\bigr)=\left\{
  \begin{array}{ll}
  {\displaystyle
  {}_4F_3\Bigl(
  \genfrac{}{}{0pt}{}{-n,\,n+\tilde{d},\,-x,\,x+d}
  {a,\,b,\,c}\Bigm|1\Bigr)}&:\text{R}\\[8pt]
  {\displaystyle
  {}_4\phi_3\Bigl(
  \genfrac{}{}{0pt}{}{q^{-n},\,\tilde{d}q^n,\,q^{-x},\,dq^x}
  {a,\,b,\,c}\Bigm|q\,;q\Bigr)}&:\text{$q$R}
  \end{array}\right.\\
  &\phantom{\check{P}_n(x;\bm{\lambda})
  =P_n\bigl(\eta(x;\bm{\lambda});\bm{\lambda}\bigr)}=\left\{
  \begin{array}{ll}
  {\displaystyle
  R_n\bigl(\eta(x;\bm{\lambda});a-1,\tilde{d}-a,c-1,d-c\bigr)}
  &:\text{R}\\[4pt]
  {\displaystyle
  R_n\bigl(\eta(x;\bm{\lambda})+1+d\,;
  aq^{-1},\tilde{d}a^{-1},cq^{-1},dc^{-1}|q\bigr)}&:\text{$q$R}
  \end{array}\right.,\nonumber
\end{align}
where $\eta(x;\bm{\lambda})$ is given by \eqref{eta} and $\tilde{d}$ is
given by \eqref{En}.
Here $R_n\bigl(x(x+\gamma+\delta+1);\alpha,\beta,\gamma,\delta\bigr)$ and
$R_n(q^{-x}+\gamma\delta q^{x+1};\alpha,\beta,\gamma,\delta|q)$ are the Racah
and $q$-Racah polynomials in the conventional parametrization \cite{kls},
respectively.
Our parametrization respects the correspondence between the ($q$-)Racah and
(Askey-)Wilson polynomials, and symmetries in $(a,b,c,d)$ are transparent.\\
$\bullet$ potential functions:
\begin{align}
  &B(x;\bm{\lambda})=
  \left\{
  \begin{array}{ll}
  {\displaystyle
  -\frac{(x+a)(x+b)(x+c)(x+d)}{(2x+d)(2x+1+d)}}&:\text{R}\\[8pt]
  {\displaystyle-\frac{(1-aq^x)(1-bq^x)(1-cq^x)(1-dq^x)}
  {(1-dq^{2x})(1-dq^{2x+1})}}&:\text{$q$R}
  \end{array}\right.,\n
  &D(x;\bm{\lambda})=
  \left\{
  \begin{array}{ll}
  {\displaystyle
  -\frac{(x+d-a)(x+d-b)(x+d-c)x}{(2x-1+d)(2x+d)}}&:\text{R}\\[8pt]
  {\displaystyle-\tilde{d}\,
  \frac{(1-a^{-1}dq^x)(1-b^{-1}dq^x)(1-c^{-1}dq^x)(1-q^x)}
  {(1-dq^{2x-1})(1-dq^{2x})}}&:\text{$q$R}
  \end{array}\right..
  \label{B,D}
\end{align}
$\bullet$ three term recurrence relations:
($P_n(\eta;\bm{\lambda})\eqdef 0$ ($n<0$))
\begin{equation}
  \eta P_n(\eta;\bm{\lambda})
  =A_n(\bm{\lambda})P_{n+1}(\eta;\bm{\lambda})
  +B_n(\bm{\lambda})P_n(\eta;\bm{\lambda})
  +C_n(\bm{\lambda})P_{n-1}(\eta;\bm{\lambda}).
  \label{3trr}
\end{equation}
$\bullet$ coefficients of the three term recurrence relations:
($A_{-1}(\bm{\lambda})\eqdef 0$)
\begin{align}
  B_n(\bm{\lambda})&=-A_n(\bm{\lambda})-C_n(\bm{\lambda}),\n
  A_n(\bm{\lambda})&=\left\{
  \begin{array}{ll}
  {\displaystyle
  \frac{(n+a)(n+b)(n+c)(n+\tilde{d})}{(2n+\tilde{d})(2n+1+\tilde{d})}}
  &:\text{R}\\[10pt]
  {\displaystyle
  \frac{(1-aq^n)(1-bq^n)(1-cq^n)(1-\tilde{d}q^n)}
  {(1-\tilde{d}q^{2n})(1-\tilde{d}q^{2n+1})}}&:\text{$q$R}
  \end{array}\right.,
  \label{AnBnCn}\\
  C_n(\bm{\lambda})&=\left\{
  \begin{array}{ll}
  {\displaystyle
  \frac{(n+\tilde{d}-a)(n+\tilde{d}-b)(n+\tilde{d}-c)n}
  {(2n-1+\tilde{d})(2n+\tilde{d})}}&:\text{R}\\[12pt]
  {\displaystyle
  d\,\frac{(1-a^{-1}\tilde{d}q^n)(1-b^{-1}\tilde{d}q^n)(1-c^{-1}\tilde{d}q^n)
  (1-q^n)}{(1-\tilde{d}q^{2n-1})(1-\tilde{d}q^{2n})}}&:\text{$q$R}
  \end{array}\right..\nonumber
\end{align}
$\bullet$ ground state eigenvector: $\phi_0(x;\bm{\lambda})>0$
\begin{equation}
  \phi_0(x;\bm{\lambda})^2=\left\{
  \begin{array}{ll}
  {\displaystyle
  \frac{(a,b,c,d)_x}{(d-a+1,d-b+1,d-c+1,1)_x}\,\frac{2x+d}{d}}
  &:\text{R}\\[12pt]
  {\displaystyle
  \frac{(a,b,c,d\,;q)_x}
  {(a^{-1}dq,b^{-1}dq,c^{-1}dq,q\,;q)_x\,\tilde{d}^x}\,\frac{1-dq^{2x}}{1-d}}
  &:\text{$q$R}
  \end{array}\right..
  \label{phi0}
\end{equation}
$\bullet$ normalization constant: $d_n(\bm{\lambda})>0$
\begin{align}
  &d_n(\bm{\lambda})^2
  =\left\{
  \begin{array}{ll}
  {\displaystyle
  \frac{(a,b,c,\tilde{d})_n}
  {(\tilde{d}-a+1,\tilde{d}-b+1,\tilde{d}-c+1,1)_n}\,
  \frac{2n+\tilde{d}}{\tilde{d}}
  }&\\[10pt]
  {\displaystyle
  \quad\times
  \frac{(-1)^N(d-a+1,d-b+1,d-c+1)_N}{(\tilde{d}+1)_N(d+1)_{2N}}
  }&:\text{R}\\[10pt]
  {\displaystyle
  \frac{(a,b,c,\tilde{d}\,;q)_n}
  {(a^{-1}\tilde{d}q,b^{-1}\tilde{d}q,c^{-1}\tilde{d}q,q\,;q)_n\,d^n}\,
  \frac{1-\tilde{d}q^{2n}}{1-\tilde{d}}
  }&\\[10pt]
  {\displaystyle
  \quad\times
  \frac{(-1)^N(a^{-1}dq,b^{-1}dq,c^{-1}dq\,;q)_N\,\tilde{d}^Nq^{\frac12N(N+1)}}
  {(\tilde{d}q\,;q)_N(dq\,;q)_{2N}}
  }&:\text{$q$R}
  \end{array}\right.\!.
  \label{dn}
\end{align}
$\bullet$ auxiliary functions:
(convention: $\prod\limits_{1\leq j<k\leq M}\!\!\!\!\!*=1$ for $M=0,1$)
\begin{align}
  \varphi(x;\bm{\lambda})&\eqdef
  \frac{\eta(x+1;\bm{\lambda})-\eta(x;\bm{\lambda})}{\eta(1;\bm{\lambda})}
  =\left\{
  \begin{array}{ll}
  {\displaystyle\frac{2x+d+1}{d+1}}&:\text{R}\\[6pt]
  {\displaystyle\frac{q^{-x}-dq^{x+1}}{1-dq}}&:\text{$q$R}
  \end{array}\right.,
  \label{varphi}\\
  \varphi_M(x;\bm{\lambda})&\eqdef\prod_{1\leq j<k\leq M}
  \frac{\eta(x+k-1;\bm{\lambda})-\eta(x+j-1;\bm{\lambda})}
  {\eta(k-j;\bm{\lambda})}\qquad
  (\varphi_0(x)=\varphi_1(x)=1)\n
  &=\prod_{1\leq j<k\leq M}
  \varphi\bigl(x+j-1;\bm{\lambda}+(k-j-1)\bm{\delta}\bigr).
  \label{varphiM}
\end{align}
$\bullet$ twist operation $\mathfrak{t}$ and twisted shift
$\tilde{\bm{\delta}}$:
\begin{equation}
  \mathfrak{t}(\bm{\lambda})\eqdef
  (\lambda_4-\lambda_1+1,\lambda_4-\lambda_2+1,\lambda_3,\lambda_4),\quad
  \tilde{\bm{\delta}}\eqdef(0,0,1,1).
\end{equation}
Note that $\eta\bigl(x;\mathfrak{t}(\bm{\lambda})\bigr)=\eta(x;\bm{\lambda})$
and $\eta(x;\bm{\lambda}+\beta\tilde{\bm{\delta}})
=\eta(x;\bm{\lambda}+\beta\bm{\delta})$ ($\beta\in\mathbb{R}$).\\
$\bullet$ virtual state polynomial $\xi_{\text{v}}(\eta;\bm{\lambda})$:
\begin{equation}
  \check{\xi}_{\text{v}}(x;\bm{\lambda})\eqdef
  \xi_{\text{v}}\bigl(\eta(x;\bm{\lambda});\bm{\lambda}\bigr)\eqdef
  \check{P}_{\text{v}}\bigl(x;\mathfrak{t}(\bm{\lambda})\bigr)
  =P_{\text{v}}\bigl(\eta(x;\bm{\lambda});
  \mathfrak{t}(\bm{\lambda})\bigr).
\end{equation}
$\bullet$ potential functions
$B'(x;\bm{\lambda})\eqdef B\bigl(x;\mathfrak{t}(\bm{\lambda})\bigr)$,
$D'(x;\bm{\lambda})\eqdef D\bigl(x;\mathfrak{t}(\bm{\lambda})\bigr)$ :
\begin{align}
  &B'(x;\bm{\lambda})=\left\{
  \begin{array}{ll}
  {\displaystyle
  -\frac{(x+d-a+1)(x+d-b+1)(x+c)(x+d)}{(2x+d)(2x+1+d)}}&:\text{R}\\[8pt]
  {\displaystyle
  -\frac{(1-a^{-1}dq^{x+1})(1-b^{-1}dq^{x+1})(1-cq^x)(1-dq^x)}
  {(1-dq^{2x})(1-dq^{2x+1})}}&:\text{$q$R}
  \end{array}\right.,\n
  &D'(x;\bm{\lambda})=\left\{
  \begin{array}{ll}
  {\displaystyle
  -\frac{(x+a-1)(x+b-1)(x+d-c)x}{(2x-1+d)(2x+d)}}&:\text{R}\\[8pt]
  {\displaystyle-\frac{cdq}{ab}\,
  \frac{(1-aq^{x-1})(1-bq^{x-1})(1-c^{-1}dq^x)(1-q^x)}
  {(1-dq^{2x-1})(1-dq^{2x})}}&:\text{$q$R}
  \end{array}\right..
  \label{B'D'}
\end{align}
$\bullet$ $\alpha(\bm{\lambda})$ and
virtual state energy $\tilde{\mathcal{E}}_{\text{v}}$:
\begin{equation}
  \alpha(\bm{\lambda})=\left\{
  \begin{array}{ll}
  1&:\text{R}\\
  abd^{-1}q^{-1}&:\text{$q$R}
  \end{array}\right.,\quad
  \tilde{\mathcal{E}}_{\text{v}}(\bm{\lambda})=\left\{
  \begin{array}{ll}
  -(c+\text{v})(\tilde{d}-c-\text{v})&:\text{R}\\[2pt]
  -(1-cq^{\text{v}})(1-c^{-1}\tilde{d}q^{-\text{v}})&:\text{$q$R}
  \end{array}\right..
  \label{Etv}
\end{equation}
$\bullet$ Casorati determinant (Casoratian) of a set of $n$ functions
$\{f_j(x)\}$ :
\begin{equation}
  \text{W}_{\text{C}}[f_1,f_2,\ldots,f_n](x)
  \eqdef\det\Bigl(f_k(x+j-1)\Bigr)_{1\leq j,k\leq n},
  \label{rdQM:Wdef}
\end{equation}
(for $n=0$, we set $\text{W}_{\text{C}}[\cdot](x)=1$).\\
$\bullet$ $r_j(x_j;\bm{\lambda},M)$ ($1\leq j\leq M+1$):
($x_j\eqdef x+j-1$)
\begin{equation}
  r_j(x_j;\bm{\lambda},M)=\left\{
  \begin{array}{ll}
  {\displaystyle
  \frac{(x+a,x+b)_{j-1}(x+d-a+j,x+d-b+j)_{M+1-j}}
  {(d-a+1,d-b+1)_M}}&:\text{R}\\[10pt]
  {\displaystyle
  \frac{(aq^x,bq^x;q)_{j-1}(a^{-1}dq^{x+j},b^{-1}dq^{x+j};q)_{M+1-j}}
  {(abd^{-1}q^{-1})^{j-1}q^{Mx}(a^{-1}dq,b^{-1}dq;q)_M}}&:\text{$q$R}
  \end{array}\right..
  \label{rj}
\end{equation}
$\bullet$ normalization constants $\mathcal{C}_{\mathcal{D}}(\bm{\lambda})$,
$\mathcal{C}_{\mathcal{D},n}(\bm{\lambda})$,
$\tilde{d}_{\mathcal{D},n}(\bm{\lambda})>0$ and
$d_{\mathcal{D},n}(\bm{\lambda})>0$ :
\begin{align}
  \mathcal{C}_{\mathcal{D}}(\bm{\lambda})&=
  \frac{1}{\varphi_M(0;\bm{\lambda})}
  \prod_{1\leq j<k\leq M}
  \frac{\tilde{\mathcal{E}}_{d_j}(\bm{\lambda})
  -\tilde{\mathcal{E}}_{d_k}(\bm{\lambda})}
  {\alpha(\bm{\lambda})B'(j-1;\bm{\lambda})},
  \label{CD}\\
  \mathcal{C}_{\mathcal{D},n}(\bm{\lambda})&=
  (-1)^M\mathcal{C}_{\mathcal{D}}(\bm{\lambda})
  \tilde{d}_{\mathcal{D},n}(\bm{\lambda})^2,
  \label{CDn}\\
  \tilde{d}_{\mathcal{D},n}(\bm{\lambda})^2&=
  \frac{\varphi_M(0;\bm{\lambda})}{\varphi_{M+1}(0;\bm{\lambda})}
  \prod_{j=1}^M\frac{\mathcal{E}_n(\bm{\lambda})
  -\tilde{\mathcal{E}}_{d_j}(\bm{\lambda})}
  {\alpha(\bm{\lambda})B'(j-1;\bm{\lambda})},
  \label{dtDn}\\
  d_{\mathcal{D},n}(\bm{\lambda})&=
  d_n(\bm{\lambda})\tilde{d}_{\mathcal{D},n}(\bm{\lambda}).
  \label{dDn}
\end{align}
$\bullet$ denominator polynomial $\Xi_{\mathcal{D}}(\eta;\bm{\lambda})$ and
multi-indexed ($q$-)Racah polynomials $P_{\mathcal{D},n}(\eta;\bm{\lambda})$:
\begin{align}
  \check{\Xi}_{\mathcal{D}}(x;\bm{\lambda})
  &\eqdef \Xi_{\mathcal{D}}\bigl(\eta(x;\bm{\lambda}+(M-1)\bm{\delta});
  \bm{\lambda}\bigr)\n
  &\eqdef\mathcal{C}_{\mathcal{D}}(\bm{\lambda})^{-1}
  \varphi_M(x;\bm{\lambda})^{-1}
  \det\bigl(\check{\xi}_{d_k}(x_j;\bm{\lambda})\bigr)_{1\leq j,k\leq M},
  \label{XiD}\\
  \check{P}_{\mathcal{D},n}(x;\bm{\lambda})
  &\eqdef P_{\mathcal{D},n}\bigl(\eta(x;\bm{\lambda}+M\bm{\delta});
  \bm{\lambda}\bigr)\n
  &\eqdef\mathcal{C}_{\mathcal{D},n}(\bm{\lambda})^{-1}
  \varphi_{M+1}(x;\bm{\lambda})^{-1}\n
  &\quad\times\left|
  \begin{array}{cccc}
  \check{\xi}_{d_1}(x_1;\bm{\lambda})&\cdots&\check{\xi}_{d_M}(x_1;\bm{\lambda})
  &r_1(x_1)\check{P}_n(x_1;\bm{\lambda})\\
  \check{\xi}_{d_1}(x_2;\bm{\lambda})&\cdots&\check{\xi}_{d_M}(x_2;\bm{\lambda})
  &r_2(x_2)\check{P}_n(x_2;\bm{\lambda})\\
  \vdots&\cdots&\vdots&\vdots\\
  \check{\xi}_{d_1}(x_{M+1};\bm{\lambda})&\cdots
  &\check{\xi}_{d_M}(x_{M+1};\bm{\lambda})
  &r_{M+1}(x_{M+1})\check{P}_n(x_{M+1};\bm{\lambda})\\
  \end{array}\right|,
  \label{PDn}
\end{align}
where $x_j\eqdef x+j-1$ and $r_j(x_j)=r_j(x_j;\bm{\lambda},M)$
($1\leq j\leq M+1$) are given in \eqref{rj}.
Other determinant expressions of $\check{P}_{\mathcal{D},n}(x;\bm{\lambda})$
can be found in \cite{detmiop}.\\
$\bullet$ coefficients of the highest degree term:
\begin{alignat*}{2}
  P_n(\eta;\bm{\lambda})
  &=c_n(\bm{\lambda})\eta^n+(\text{lower order terms}),
  &\ \ P_{\mathcal{D}}(\eta;\bm{\lambda})
  &=c_{\mathcal{D},n}^{P}(\bm{\lambda})\eta^{\ell_{\mathcal{D}}+n}
  +(\text{lower order terms}),\\
  \xi_{\text{v}}(\eta;\bm{\lambda})
  &=\tilde{c}_{\text{v}}(\bm{\lambda})\eta^{\text{v}}
  +(\text{lower order terms}),
  &\ \ \Xi_{\mathcal{D}}(\eta;\bm{\lambda})
  &=c_{\mathcal{D}}^{\Xi}(\bm{\lambda})\eta^{\ell_{\mathcal{D}}}
  +(\text{lower order terms}),
\end{alignat*}
\begin{align}
  c_n(\bm{\lambda})&=\left\{
  \begin{array}{ll}
  {\displaystyle\frac{(\tilde{d}+n)_n}{(a,b,c)_n}}&:\text{R}\\[8pt]
  {\displaystyle\frac{(\tilde{d}q^n;q)_n}{(a,b,c;q)_n}}&:\text{$q$R}
  \end{array}\right.,\quad
  \tilde{c}_{\text{v}}(\bm{\lambda})=\left\{
  \begin{array}{ll}
  {\displaystyle\frac{(c+d-a-b+\text{v}+1)_{\text{v}}}
  {(d-a+1,d-b+1,c)_{\text{v}}}}&:\text{R}\\[8pt]
  {\displaystyle\frac{(a^{-1}b^{-1}cdq^{\text{v}+1};q)_{\text{v}}}
  {(a^{-1}dq,b^{-1}dq,c;q)_{\text{v}}}}&:\text{$q$R}
  \end{array}\right.,\\
  c_{\mathcal{D}}^{\Xi}(\bm{\lambda})&=
  \prod_{j=1}^M\tilde{c}_{d_j}(\bm{\lambda})\times\left\{
  \begin{array}{ll}
  {\displaystyle
  \frac{\prod_{j=1}^M(d-a+1,d-b+1,c)_{j-1}}
  {\prod\limits_{1\leq j<k\leq M}(c+d-a-b+d_j+d_k+1)}}
  &:\text{R}\\[22pt]
  {\displaystyle
  \frac{\prod_{j=1}^M(a^{-1}dq,b^{-1}dq,c;q)_{j-1}}
  {\prod\limits_{1\leq j<k\leq M}(1-a^{-1}b^{-1}cdq^{d_j+d_k+1})}}
  &:\text{$q$R}
  \end{array}\right.,
  \label{cXiD}\\
  c_{\mathcal{D},n}^{P}(\bm{\lambda})&=
  c_{\mathcal{D}}^{\Xi}(\bm{\lambda})c_n(\bm{\lambda})
  \times\left\{
  \begin{array}{ll}
  {\displaystyle\prod_{j=1}^M\frac{c+j-1}{c+d_j+n}}&:\text{R}\\[6pt]
  {\displaystyle\prod_{j=1}^M\frac{1-cq^{j-1}}{1-cq^{d_j+n}}}&:\text{$q$R}
  \end{array}\right..
  \label{cPDn}
\end{align}
$\bullet$ coefficients $g_n^{\prime\,(k)}(\bm{\lambda})$:
\begin{equation}
  \frac{\eta(x;\bm{\lambda})^{n+1}-\eta(x-1;\bm{\lambda})^{n+1}}
  {\eta(x;\bm{\lambda})-\eta(x-1;\bm{\lambda})}
  =\sum_{k=0}^ng_n^{\prime\,(k)}(\bm{\lambda})
  \eta(x;\bm{\lambda}-\bm{\delta})^{n-k}
  \ \ (n\in\mathbb{Z}_{\geq 0}),
\end{equation}
where $g_n^{\prime\,(k)}(\bm{\lambda})$ is given by
\begin{align}
  \text{R}:\ \ &g_n^{\prime\,(k)}(\bm{\lambda})\eqdef
  \sum_{r=0}^k\sum_{l=0}^{k-r}
  \genfrac{(}{)}{0pt}{}{n+1}{r}\genfrac{(}{)}{0pt}{}{n-r-l}{n-k}
  (-1)^{r+l}\Bigl(\frac{d}{2}\Bigr)^{2r}
  \Bigl(\frac{d-1}{2}\Bigr)^{2(k-r-l)}
  g_{n-r}^{\prime\,(l)\,\text{W}},\\
  \text{$q$R}:\ \ &g_n^{\prime\,(k)}(\bm{\lambda})\eqdef
  \sum_{r=0}^k\sum_{l=0}^{k-r}
  \genfrac{(}{)}{0pt}{}{n+1}{r}\genfrac{(}{)}{0pt}{}{n-r-l}{n-k}
  (-1)^r\bigl(2d^{\frac12}\bigr)^lq^{\frac12(n-r-l)}
  \bigl(1+d\bigr)^r\bigl(1+dq^{-1}\bigr)^{k-r-l}\n
  &\phantom{g_n^{\prime\,(k)}(\bm{\lambda})\eqdef}\times
  g_{n-r}^{\prime\,(l)\,\text{AW}},
\end{align}
Here $g_n^{\prime\,(k)\,\text{W}}$ and $g_n^{\prime\,(k)\,\text{AW}}$ are
\cite{rrmiop2}
\begin{align*}
  &g_n^{\prime\,(k)\,\text{W}}\eqdef
  \frac{(-1)^k}{2^{2k+1}}\genfrac{(}{)}{0pt}{}{2n+2}{2k+1},\n
  &g_n^{\prime\,(k)\,\text{AW}}\eqdef
  \theta(\text{$k$\,:\,even})\frac{(n+1)!}{2^k}
  \sum_{r=0}^{\frac{k}{2}}\genfrac{(}{)}{0pt}{}{n-k+r}{r}
  \frac{(-1)^rq^{-\frac12(n-k+2r)}}{(\frac{k}{2}-r)!\,(n-\frac{k}{2}+1+r)!}
  \frac{1-q^{n-k+1+2r}}{1-q},
\end{align*}
and $\theta(P)$ is a step function for a proposition $P$,
$\theta(P)=1$ ($P$ : true), $0$ ($P$ : false).\\
$\bullet$ map $I_{\bm{\lambda}}:\{\text{polynomial}\}\to\{\text{polynomial}\}$:
\begin{equation}
  p(\eta)=\sum_{k=0}^na_k\eta^k\mapsto
  I_{\bm{\lambda}}[p](\eta)\eqdef\sum_{k=0}^{n+1}b_k\eta^k,
  \label{mapI}
\end{equation}
where $b_k$'s are defined by
\begin{equation}
  b_{k+1}=\frac{1}{g_k^{\prime\,(0)}(\bm{\lambda})}
  \Bigl(a_k-\sum_{j=k+1}^ng_j^{\prime\,(j-k)}(\bm{\lambda})b_{j+1}\Bigr)
  \ \ (k=n,n-1,\ldots,1,0),\quad
  b_0=0.
\end{equation}


\end{document}